\documentclass[aps,pre,superscriptaddress,twocolumn,balancelastpage]{revtex4-2}

\usepackage[colorlinks,bookmarks=false,citecolor=blue,linkcolor=blue,urlcolor=blue]{hyperref}
\usepackage[all]{hypcap}   

\usepackage{amsmath,amssymb}
\usepackage{graphicx}
\usepackage{enumerate}

\usepackage{verbatim}
\usepackage{color}

\usepackage{placeins}  

\usepackage{braket}
\usepackage{dsfont}

\usepackage{physics}

\DeclareMathOperator\sinc{sinc}

\DeclareMathOperator\LambW{W}
\DeclareMathOperator\Wg{Wg}

\usepackage{enumitem}

\newcommand{\ue}{\text{e}}
\newcommand{\ui}{\text{i}}
\newcommand{\ud}{\text{d}}
\newcommand{\ua}{\text{A}}
\newcommand{\ub}{\text{B}}
\newcommand{\uc}{\text{c}}

\newcommand{\U}{\mathcal{U}}

\newcommand{\HIDDEN}[1]{}

\makeatletter
\let\Hy@backout\@gobble
\makeatother

\begin{document}

\title{Universal spectral correlations in interacting chaotic few-body quantum systems}

\author{Felix Fritzsch}
\affiliation{Physics Department, Faculty of Mathematics and Physics,
    University of Ljubljana, Ljubljana, Slovenia}
\affiliation{Max Planck Institute for the Physics of Complex Systems, Dresden, Germany}

\author{Maximilian~F.~I.~Kieler}
\affiliation{Technische Universit\"at Dresden,
 Institut f\"ur Theoretische Physik and Center for Dynamics,
 Dresden, Germany}

\date{\today}

\begin{abstract}
    The emergence of random matrix spectral correlations in interacting quantum 
    systems is a defining feature of quantum chaos.    
    We study such correlations in terms of the spectral form factor and its moments in interacting chaotic few- and
    many-body systems, modeled by suitable random-matrix ensembles.
    We obtain the spectral form factor exactly for large Hilbert space dimension.
    Extrapolating those results to finite Hilbert space dimension we find a 
    universal transition from the non-interacting to the strongly interacting case, which can be described as a simple combination of these two limits.
    This transition is governed by a single scaling parameter.
    In the bipartite case we derive similar results also for all moments of the spectral form factor.
    We confirm our results by extensive numerical studies and demonstrate that they apply to more realistic systems given by a pair of quantized kicked rotors as well.
    Ultimately we complement our analysis by a perturbative approach covering the small 
    coupling regime.
\end{abstract}

\maketitle

\section{Introduction}

Quantizing a classically chaotic systems leads to energy spectra whose fluctuations universally coincide with those of random matrices \cite{CasValGua1980,Ber1981b,BohGiaSch1984} which depend only on the symmetries of the system at hand \cite{Dys1962b,Wig1967,Meh1991,Haa2010,Sto2000}.
This universality can be traced back to only a few basic properties of the underlying ergodic dynamical system \cite{Gut1990,SieRic2001,Sie2002,MueHeuBraHaaAlt2004}.
Random-matrix like spectral fluctuations henceforth have become one of the most widely accepted definitions of quantum chaos even in the absence of a classical limit.
Spectral fluctuations are often studied numerically on the scale of the mean level spacing in terms of nearest neighbor level spacings.
In contrast, spectral correlations on all energy scales are conveniently probed by the 
spectral form factor \cite{Haa2010}, which has received a considerable amount of attention in recent years in, e.g., high-energy physics \cite{CotHunLiuYos2017,CotGurHanPolSaaSheStaStrTez2017,GhaHanSheTez2018,WinSwi2022} and  condensed-matter and many-body systems \cite{KosLjuPro2018,BerKosPro2018,BerKosPro2021,CheLud2018,SunBonProVid2020,ChaDeCha2018b,ChaDeCha2018,FriChaDeCha2019,GarCha2021,ChaDeCha2021,BerKosPro2022,DagMisChaSad2022:p,WinBarBalGalSwi2022a,BarWinBalSwiGal2023:p,FlaBerPro2020,KosBerPro2021,MouPreHusCha2021,AkiWalGutGuh2016}.
In particular for the latter, random-matrix spectral correlations have to arise from the mutual -- often local -- interactions of the individual subsystems, e.g., particles, spins, etc.
To date, there is no general universal underlying mechanism for this emergent behavior comparable to the semiclassical periodic orbit based picture in single-particle systems.
However, there are a few examples, in which exact results are obtained.
This includes random quantum circuits with large local Hilbert space dimension $N \to \infty$  \cite{ChaDeCha2018b,ChaDeCha2018,GarCha2021,ChaDeCha2021} or spatially homogeneous circuits or Floquet spin chains in the thermodynamic limit of system size $L \to \infty$ \cite{KosLjuPro2018,BerKosPro2018,BerKosPro2021,BerKosPro2022}.
It nevertheless remains of great interest to identify minimal models in which spectral correlations, i.e., the spectral form factor, can be obtained accurately even for finite $N$ and $L$ to reveal how random-matrix behavior emerges.

One of the arguably simplest scenarios is that of just two subsystems being coupled with each other.
When both subsystems are individually chaotic, the level spacing distribution shows a transition from Poissonian statistics in the uncoupled case towards Wigner-Dyson statistics in the strong coupling regime following a universal scaling law \cite{SriTomLakKetBae2016}.
The latter can be explained in a suitable random matrix model adapted to the bipartite setting -- the so-called random matrix transition ensemble (RMTE) \cite{SriTomLakKetBae2016}.
Subsequently a universal transition was observed also in the eigenstates reflected by, e.g., their entanglement and localization properties \cite{LakSriKetBaeTom2016,TomLakSriBae2018,HerKieFriBae2020} as well as in the entanglement dynamics after a quench \cite{PulLakSriBaeTom2020,PulLakSriKieBaeTom2023}.
\\

In this work we analyze the spectral form factor in both the bipartite RMTE as well as in an extended version thereof.
The latter models interacting few- and many-body systems consisting of $L$ subsystems of size $N$ subject to an all-to-all interaction.
We derive exact results for the spectral form factor and, in the bipartite case, its moments as
the size of the subsystems $N \to \infty$ for an arbitrary number of subsystems $L$.
These exact results naturally extend to large, but finite, subsystem size $N$ and reveal a universal dependence of the spectral form factor on a single scaling parameter, which fully captures the influence of $N$, $L$, and the nature of the interaction between the subsystems.
This gives rise to a universal transition of the spectral form factor from the non-interacting to the strongly interacting regime.
In fact, the spectral form factor can be written as a simple time dependent convex combination of these two limiting cases.
The spectral form factor consequently exhibits an intricate interplay between different time and associated energy scales, including the Heisenberg times of the subsystems and the full system as well as a non-trivial Thouless time.

We begin by deriving our results in the simplest setting of the bipartite RMTE, which already shows most of the phenomena indicated above.
We focus on systems which lack time-reversal or other antiunitary symmetries and model the subsystems by random matrices from the circular unitary ensemble $\text{CUE}(N)$ of dimension $N$.
We couple them via  a random $N^2$ dimensional diagonal matrix with tunable coupling, i.e., interaction, strength.
Within this framework we derive exact expressions of the spectral form factor and all its moments for initial times by explicitly computing the Haar averages over the subsystems.
Subsequently we show a natural way to extend these results to larger times. This allows for expressing the spectral form factor as a time dependent convex combinations of the spectral form factor of two uncoupled $\text{CUE}(N)$ and that of $\text{CUE}(N^2)$.
We extract a single scaling parameter $\Gamma$ from this expression which universally governs the transition for times larger than the subsystems Heisenberg times
from the former uncoupled, non-interacting result at $\Gamma=0$ towards the later full random matrix result at large $\Gamma$.
For intermediate scaling parameter $\Gamma$ we extract the Thouless time as the time scale after which the spectral form factor of the RMTE agrees with that of $\text{CUE}(N^2)$.
For the fluctuations of the spectral form factor expressed in terms of its higher moments we
find a similar expression as they can be written as time dependent convex combinations of products of moments of the spectral form factor of two uncoupled $\text{CUE}(N)$ and $\text{CUE}(N^2)$.
Again, for large enough times the moments depend exclusively on the scaling parameter $\Gamma$ and thus exhibit a similar universal transition.
We confirm all our analytical results by comparison with extensive numerical studies.

To verify, that the above results derived in a random matrix model apply to physical models as well, we study the spectral form factor and its moments in two coupled quantum kicked rotors in the chaotic regime.
There, we find good agreement between the RMTE predictions and the numerically computed spectral form factor and its moments.
This includes in particular the universal dependence on a single scaling parameter.
Moreover, we argue, that for moderately large scaling parameter Thouless time and Ehrenfest time, i.e., the time up to which quantum dynamics follows classical dynamics, do not coincide.

Having established the methods and results for the bipartite case, we eventually introduce a many-body version of the bipartite RMTE.
It is built from $L$ independent $\text{CUE}(N)$ matrices modeling the subsystems, whereas an all-to-all interaction of tunable strength is induced by an $N^L$ dimensional random diagonal unitary matrix.
The results obtained for the bipartite case directly carry over to this many-body setting.
In particular the spectral form factor is a time dependent convex combination of the spectral form factor of $L$ non-interacting $\text{CUE}(N)$ and of the full $\text{CUE}\left(N^L\right)$, which at times larger than the subsystems Heisenberg time depends on a single scaling parameter $\Gamma$ only. 
As in the bipartite case, we verify those results by extensive numerical studies.
Even though those are limited to the few-body setting of small $L$ we expect our results to apply also in the many-body case of large $L$.

Ultimately, we complement our approach with a properly regularized perturbative treatment of the coupling, which captures the regime of small scaling parameter and accurately describes the spectral form factor for large times way beyond the Heisenberg time of the full system both for the bipartite and the extended case. \\

The remainder of this paper is organized as follows. In Sec.~\ref{sec:rmte} we review the bipartite RMTE, whereas 
Sec.~\ref{sec:sff_intro} gives a short introduction into the spectral form factor as the main object of our work. Subsequently we derive the spectral form factor for large $N$ in Sec~\ref{sec:sff_semiclassics} and establish its universal dependence on a single scaling parameter in Sec.~\ref{sec:universality}.
This allows for computing the Thouless time in Sec.~\ref{sec:thouless_time}.
We then proceed by deriving the higher moments of the spectral form factor in 
Sec.~\ref{sec:sff_moments} and apply our results to a system of coupled kicked 
rotors in Sec.~\ref{sec:kicked_rotors}.
In Sec.~\ref{sec:ext_rmte} we discuss the extended version of the RMTE.
The perturbative treatment of the coupling is then presented in Sec.~\ref{sec:perturbation_theory}.
We finally summarize our results in Sec.~\ref{sec:conclusions}.

\section{Random Matrix Transition Ensemble}
\label{sec:rmte}

In this section we review the random matrix transition ensemble (RMTE) introduced in Ref.~\cite{SriTomLakKetBae2016}, which allows for studying universal features of coupled bipartite chaotic quantum systems. 
We consider ensembles of Floquet systems evolving in discrete time steps with 
evolution between subsequent time steps governed by a unitary evolution 
operator $\U \in \text{U}(N^2)$.
The model is built from individual subsystems $\ua$ and $\ub$ described by 
Hilbert spaces $\mathcal{H}_\ua \simeq \mathcal{H}_\ub \simeq \mathds{C}^N$ of 
dimension $N$.
Note, that an extension to subsystem Hilbert spaces 
with different dimensions is straight forward \cite{HerKieFriBae2020}.
The dynamics of the individual subsystems is governed by unitary evolution 
operators $\U_\ua, \U_\ub \in \text{U}(N)$ and the coupling is modeled by a 
unitary $\U_{\uc}(\epsilon) \in \text{U}(N^2)$.
Here $\epsilon$ governs the strength of the coupling with $\epsilon=0$ 
corresponding to the uncoupled situation $\U_\uc(0) =\mathds{1}_{N^2}$.
The coupled bipartite system is described by the Hilbert space 
$\mathcal{H} = \mathcal{H}_\ua \otimes \mathcal{H}_\ub \simeq \mathds{C}^{N^2}$
We denote the canonical basis for the subsystems by $\ket{i}$ and the corresponding product basis in $\mathcal{H}$ by $\ket{ij}$, $i, j \in \{1,\ldots, N\}$.

The Floquet operator for the coupled system then reads
\begin{align}
    \U = \U_\uc(\epsilon)\left(\U_\ua \otimes \U_\ub \right).
    \label{eq:rmte}
\end{align}
We focus on systems obeying no anti-unitary symmetry, e.g., time-reversal 
symmetry and hence choose $\U_A$ and $\U_B$ independently from the circular 
unitary ensemble CUE($N$), i.e., Haar-random from  $\text{U}(N)$.
The coupling is modeled by a diagonal matrix with matrix elements
\begin{align}
    \bra{ij}\U_{\uc}(\epsilon)\ket{mn} =  
    \delta_{im}\delta_{jn} \exp(\ui \epsilon \xi_{ij})
    \label{eq:coupling}
\end{align}
with i.i.d. random phases $\xi_{ij}$, $i,j \in \{1,\ldots, N\}$ with arbitrary 
distribution with finite first and second moment.
The first moment, i.e., the expectation value $\langle \xi \rangle_\xi$, where $\langle \cdot \rangle_\xi$ denotes the average with respect to the distribution of the phases, merely induces an overall phase for the Floquet operator $\U$.
We hence might assume it to be zero in the following.
In contrast the second moment, i.e., the variance
\begin{align}
    \sigma^2 = 
    \big\langle \xi^2 \big\rangle_\xi,
\end{align}
as well es the real and positive coupling strength $\epsilon$
give rise to an effective coupling strength $\sigma\epsilon$.
In principle, both parameters could be combined into a single one, but for later convenience we keep both $\sigma$ and $\epsilon$.
The latter governs the strength of the coupling once the distribution of the phases $\xi_{ij}$, and hence $\sigma$, is fixed.
We refer to $\epsilon=0$ as the uncoupled case as $\U_\uc(0)=\mathds{1}$ and to large $\epsilon \sim 1$ as the strongly coupled case.

For a fixed realization of $\U$ from the RMTE the eigenvalues and eigenvectors of $\U$ obey
\begin{align}
\U \ket{\varphi_{ij}} = \ue^{\ui \varphi_{ij}} \ket{\varphi_{ij}}
\label{eq:eigenvalues}
\end{align}
with eigenphases (quasi energies) $\varphi_{ij} \in \left[-\pi, \pi\right)$.
Here we index eigenphases by double indices $ij$, with $i,j \in \{1,\ldots, N\}$,  for later convenience, when treating the eigenphases perturbatively in $\epsilon$.
This is motivated by the uncoupled case $\epsilon=0$ for which the Floquet operator 
$\U$ is a tensor product and hence the eigenvectors are products 
$\ket{\vartheta_i^\ua \vartheta_j^\ub}=\ket{\vartheta_i^\ua}\otimes\ket{
\vartheta_j^\ub } $ of eigenvectors of $\U_\ua$ and $\U_\ub$, with eigenphases 
$\vartheta_i^\ua$ and $\vartheta_j^\ub$, respectively.
Consequently the eigenphases of $\U$ are of the form $\varphi_{ij}(\epsilon=0) 
= \vartheta_i^\ua + \vartheta_j^\ub \! \mod \! 2 \pi$ and are uniformly distributed in $[-\pi , \pi)$.
In the remainder of the paper we consider arithmetic operations on eigenphases modulo $2 \pi$ but suppress it in the notation.
Our main focus is on statistical properties of the eigenphases of the coupled system as a function of both $N$ and $\epsilon$. 

\section{Spectral Form Factor}
\label{sec:sff_intro}

In this section we briefly review some basic properties of the statistics of 
eigenphases and in particular of the spectral form factor as a measure for 
correlations in the (quasi) energy spectrum.
For members of the RMTE introduced in the previous section the spectral density reads
\begin{align}
    \rho(\varphi) = \frac{2\pi}{N^2} \sum_{i,j=1}^{N}\delta(\varphi - 
\varphi_{ij}).
    \label{eq:spectral_density}
\end{align}
Here, the normalization is chosen such that the mean spectral density
\begin{align}
\langle \rho(\varphi) \rangle_\varphi = \frac{1}{2\pi}\int_{-\pi}^\pi \rho(\varphi) \ud  \varphi = 1
\label{eq:spectral_density2}
\end{align}
is unity.
Correlations in the spectrum can then be described by the connected two-point correlation function
\begin{align}
r(\omega) &  = \frac{1}{2\pi}\int_{-\pi}^\pi \rho(\varphi - 
\omega/2)\rho(\varphi + \omega/2)\ud \varphi - \langle \rho(\varphi) 
\rangle_\varphi^2
\label{eq:spectral_twopoint_function} \\
& = \left(\frac{2\pi}{N^4}\sum_{i,j, k,l=0}^{N}\delta(\varphi_{ij} - \varphi_{kl} - \omega)\right)   -  1.  \label{eq:spectral_twopoint_function2}
\end{align}
The ensemble average of its Fourier transform finally defines the spectral form factor as a function of discrete time $t$ as
\begin{align}
K(t) & = \frac{N^4}{2\pi}\Big\langle \int_{-\pi}^\pi \ue^{\ui \omega t}  r(\omega) \ud \omega \Big\rangle \\
& = \Big\langle \sum_{i,j, k,l=0}^{N}\ue^{\ui (\varphi_{ij} - \varphi_{kl})t}\Big\rangle  - N^4\delta_{t0}.
\label{eq:spectral_form_factor}
\end{align}
Here the bracket denotes the average over the RMTE, i.e., the Haar averages 
over $\U_\ua$ and $\U_\ub$ as well as over the random phases 
$\xi_{ij}$.
This averaging procedure is necessary as the spectral form factor is not self averaging \cite{Pra1997} but fluctuates wildly for a single realization.
The spectral form factor is particularly convenient to study as Eq.~\eqref{eq:spectral_form_factor} can be written in terms of the Floquet operator as
\begin{align}
K(t) = \big\langle | \text{tr}\left(\U^t\right) | ^2 \big\rangle  - N^4\delta_{t0}.
\label{eq:spectral_form_factor_trace}
\end{align}
For Floquet operators $\U$ drawn not from the RMTE but from CUE($M$) (in our case $M=N$ or $M=N^2$) the spectral form factor reads \cite{Haa2010}
\begin{align}
K_M(t) = K_{\text{CUE}(M)}(t) = \min\{t, M\}
\label{eq:spectral_form_factor_CUE}
\end{align}
and is characterized by a linear ramp $K(t)=t$ up to $t=M$ and a subsequent plateau $K(t)=M$ for times $t\geq M$. 
Note that $t=t_{\text{H}}=M$ corresponds to the Heisenberg time determined by the inverse mean level spacing $2 \pi /M$.

\begin{figure}[]
    \centering
    \includegraphics[width=8.5cm]{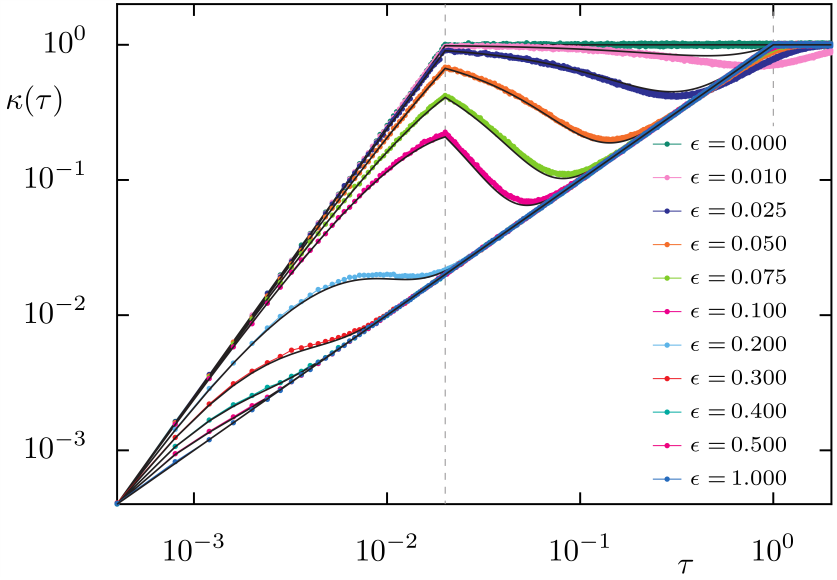}
    \caption{Rescaled spectral form factor $\kappa(\tau)$ for the RMTE at $N=50$ 
and different coupling strengths $\epsilon$ (see legend, increasing from top to bottom) in log-log scale. 
Colored symbols correspond to numerical data obtained from $20000$ realizations 
of the RMTE with uniformly distributed phases $\xi_{ij}$. The asymptotic 
result~\eqref{eq:spectral_form_factor_semiclassics2} is depicted as black lines.
        Dashed gray lines correspond to the Heisenberg time $\tau_{\text{SH}}$ of the subsystems and of the bipartite system $\tau_{\text{H}}$.}
    \label{fig:sff_rmte_N50}
\end{figure}

\begin{figure}[]
    \centering
    \includegraphics[width=8.5cm]{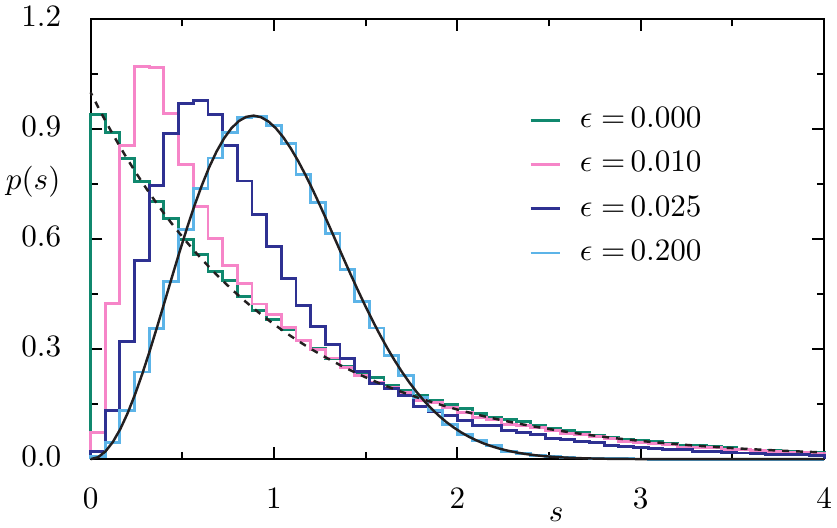}
    \caption{Level spacing distribution for the RMTE at $N=50$ and different 
coupling strengths $\epsilon$ (see legend, increasing from left to right). Colored histograms  correspond to 
numerical data obtained from 100 realizations of the RMTE with uniformly 
distributed phases $\xi_{ij}$. The solid black line represents the random matrix 
result for the CUE, Eq.~\eqref{eq:wigner_surmise}, whereas the dashed black line represents Poissonian 
statistics.}
    \label{fig:level_spacing_rmte_N50}
\end{figure}

For the RMTE we compute the spectral form factor numerically for $N=50$ and various coupling strengths $\epsilon \in [0, 1]$.
We choose the phases $\xi_{ij}$ entering $\U_{\uc}$ to be uniformly 
distributed in $[-\pi, \pi]$ and average over more than $20000$ 
realizations.
The resulting spectral form factors are depicted in Fig.~\ref{fig:sff_rmte_N50} as colored symbols connected by lines.
There we also introduce both the rescaled spectral form factor 
\begin{equation}
 \kappa(\tau)=K(t)/N^2
 \label{eq:rescaled_spectral_form_factor}
\end{equation}
as well as rescaled time $\tau = t/N^2$, i.e., both are scaled by the 
Heisenberg time $t_{\text{H}}=N^2$ of the bipartite system.
In this units the Heisenberg time of the full system reads $\tau_{\text{H}}=1$ , whereas those of the subsystems read $\tau_{\text{SH}}=1/N$.
For the uncoupled case $\epsilon=0$, the evolution operator is the tensor 
product of two independent CUE($N$) matrices each having spectral form factor 
given by Eq.~\eqref{eq:spectral_form_factor_CUE}.
Using the multiplicativity of the trace with respect to tensor products, $\text{tr}\left(U_{\text{A}}\otimes U_{\text{B}}\right)=\text{tr}\left(U_{\text{A}}\right) \text{tr}\left(U_{\text{B}}\right)$, and the representation~\eqref{eq:spectral_form_factor_trace}, the spectral form factor of the uncoupled system factorizes, see also Ref.~\cite{AgrPan2021}.
Consequently the spectral form factor of the uncoupled bipartite system is given by $K(t)=t^2$ for $t\leq N$ and $K(t)=N^2$ for $t\geq N$.
That is, for times up to the Heisenberg time of the subsystems 
$t_{\text{SH}}=N$ the spectral form factor grows 
quadratically and reaches a plateau at later times with the plateau value given 
by $N^2 = \dim \mathcal{H}$.
This is well confirmed by the numerical data shown in 
Fig.~\ref{fig:sff_rmte_N50}.
In contrast, for strong coupling the spectral form factor of the RMTE is expected to reproduce the random matrix result for CUE($N^2$), Eq.~\eqref{eq:spectral_form_factor_CUE}, as it is indeed the case for $\epsilon=1$.
In the intermediate coupling regime up to a time scale $t_{\text{Th}}$ --- the 
so-called Thouless time --- we observe non-universal behavior, in the sense, 
that $K(t)$ does not follow the random matrix 
result~\eqref{eq:spectral_form_factor_CUE} for CUE$(N^2)$.
For times $t>t_{\text{Th}}$ the spectral form factor follows the linear growth of  $K_{N^2}(t)$ and the subsequent plateau.
Moreover, in this regime, the transition from the ramp regime to the plateau is not sharp but smoothed out around $t=t_{\text{H}}$ ($\tau=1$) as a consequence of the sum rule
$\sum_t (K(t) - K_{N^2}(t))=0$ \cite{Haa2010} which requires compensation of the non-universal initial regime.
An additional interesting feature is the sharp transition from increasing 
to decreasing spectral form factor at time $t=t_{\text{SH}}$ ($\tau=1/N$).

It is instructive to compare the transition of the spectral form factor from the 
uncoupled to the strongly coupled case with the corresponding transition of the 
level spacing distribution $p(s)$ where $s$ is the spacing between subsequent 
normalized eigenphases, $s=\frac{N^2}{2\pi} \left( \varphi_{i+1} - \varphi_{i} 
\right)$ after arranging the eigenphases in increasing order.
For the tensor product of two uncoupled CUE($N$) matrices the eigenphases of the bipartite system are an uncorrelated superposition of the individual spectra and follow Poissonian statistics as $N \to \infty$ \cite{TkoSmaKusZeiZyc2012}.
Hence the distribution of level spacings is exponential, $p(s)=\exp(-s)$.
In contrast for the strongly coupled case we expect the level spacing distribution to be well described by the Wigner surmise for CUE given by
\begin{align}
    p(s) = \frac{32}{\pi^2}s^2 \ue ^{-\frac{4}{\pi}s^2}.
    \label{eq:wigner_surmise}
\end{align}
We depict the numerically obtained level spacing distribution for the same RMTE as used for Fig.~\ref{fig:sff_rmte_N50} in Fig.~\ref{fig:level_spacing_rmte_N50} for representative values of the coupling strength.
We observe a transition of the level spacing distribution from the Poissonian statistics at $\epsilon=0$ to the distribution~\eqref{eq:wigner_surmise} at $\epsilon=0.2$.
For larger coupling strength the level spacing distribution is not shown as it coincides with $\epsilon=0.2$ and Eq.~\eqref{eq:wigner_surmise}.
The transition towards full CUE random matrix statistics is much faster for the
level spacing distribution than for the spectral form factor, which at short times shows significant deviations from the random matrix result~\eqref{eq:spectral_form_factor_CUE}.
On the one hand this is not unexpected, as the level spacing distribution describes spectral correlations on the scale of the mean level spacing, whereas the spectral form factor probes correlations at all energy scales.
In particular correlations at the scale of the mean level spacing are probed at late times $t \approx t_{\text{H}}$.
In this time regime, the spectral form factor approaches the random matrix result~\eqref{eq:spectral_form_factor_CUE} already for coupling strengths, for which also the level spacing distribution follows random matrix theory.
On the other hand we might be lead to the conclusion that the spectral form factor is a more sensitive probe for spectral correlations in dependence of the coupling strength.
This is the case for correlations on larger energy scales and hence
governs the short time properties, e.g. the relaxation dynamics towards equilibrium.
In contrast long time and steady state properties corresponding to spectral correlations on small energy scales are less sensitive to the coupling.

\section{Exact Spectral Form Factor in the Semiclassical Limit 
\label{sec:sff_semiclassics}}

In the following section we provide a qualitatively accurate description of the 
spectral form factor by deriving its asymptotics for large $N$.
As $1/N$ plays the role of an effective Planck's constant we refer to $N\to \infty$ as the semiclassical limit.
We essentially follow the derivation presented in Ref.~\cite{ChaDeCha2018,ChaDeCha2018b}, where a 
spatially extended version of the RMTE in the form of a random quantum circuit 
was presented.
Our objective is to evaluate the Haar average over the subsystems $\U_\ua$ and 
$\U_\ub$ exactly in the limit $N \to \infty$ first and subsequently average over the
random phases in the coupling $\U_{\text{c}}$.
To this end we introduce some useful notation in the following.
For $t>0$ we denote the product basis in $\mathcal{H}_\ua^{\otimes t}$ by 
$\ket{\mathbf{i}}=\ket{i_1,i_2,\ldots,i_t}$ with $i_s \in \{1,\ldots,N\}$ and 
similar for the product basis in $\mathcal{H}_\ub^{\otimes t}$.
For the corresponding product basis in $\mathcal{H}^{\otimes t} = 
\left(\mathcal{H}_\ua \otimes\mathcal{H}_\ub\right)^{\otimes t}$, after a 
suitable rearrangement of tensor factors, we write 
$\ket{\mathbf{i},\mathbf{j}}=\ket{i_1, j_1}\otimes\cdots\otimes\ket{i_t,j_t}$.
Regarding the expansion in time, we consider the action of the permutation 
group $S_t$ of $t$ elements on 
$\mathcal{H}_\ua^{\otimes t}$ which permutes tensor factors and denote the 
action of $\pi \in S_t$ on the above product basis by
\begin{align}
    \ket{\pi(\mathbf{i})}=\ket{i_{\pi^{-1}(1)},\ldots,i_{\pi^{-1}(t)}}.
\end{align}
A central role is played by the $t$-periodic shifts $\eta_r \in S_t$, defined 
by 
$\eta_r(s)=s+r \mod t$ for $s \in \{1,\ldots,t\}$ and $r \in \{0,\ldots t-1\}.$

To proceed, we rewrite Eq.~\eqref{eq:spectral_form_factor_trace} for $t>0$ as
\begin{align}
K(t)  = \text{tr}\left(\U^t\right)\tr\left(\left[\U^\star\right]^t\right)
\end{align}
where $\U^{\star}$ denotes the complex conjugate matrix of $\U$ and we use the invariance of the trace under taking the transpose.
Using the above notation this can be expressed as
\begin{align}
    K(t) & = \Big\langle \sum_{\mathbf{i},\mathbf{j}}
    \bra{\mathbf{i},\mathbf{j}}\U^{\otimes t}\ket{\eta_1(\mathbf{i}),\eta_1(\mathbf{j})} \nonumber \\
    & \qquad  \times \sum_{\mathbf{k},\mathbf{l}}
    \bra{\mathbf{k},\mathbf{l}}\left(\U^\star \right)^{\otimes t}\ket{\eta_1(\mathbf{k}),\eta_1(\mathbf{l})} \Big\rangle.
    \label{eq:spectral_form_factor_tensor_products}
\end{align}
Here the fourfold sum runs over all elements of the product basis introduced 
above.
The above expression is essentially obtained by introducing a suitable number of 
resolutions of identity in terms of the basis states $\ket{i j} \in \mathcal{H}$ in the $t$-fold products $\U^t$ and $\left[U^{\star}\right]^t$. 

For the matrix elements of $\U^{\otimes t}$ we obtain by inserting Eqs.~\eqref{eq:rmte}~and~\eqref{eq:coupling} 
\begin{align}
    \bra{\mathbf{i},\mathbf{j}}\U^{\otimes t}\ket{\eta_1(\mathbf{i}),\eta_1(\mathbf{j})} =  
    \qquad \qquad \qquad \qquad  \qquad \qquad \qquad \qquad  \nonumber \\
    \exp\left(\ui \epsilon \sum_{s=1}^t \xi_{i_s j_s}\right)
   \bra{\mathbf{i}}\U_\ua^{\otimes t}\ket{\eta_1(\mathbf{i})}
   \bra{\mathbf{j}}\U_\ub^{\otimes t}\ket{\eta_1(\mathbf{j})}, \hspace*{1.95cm}\\
    \bra{\mathbf{k},\mathbf{l}}\left(\U^\star \right)^{\otimes t}\ket{\eta_1(\mathbf{k}),\eta_1(\mathbf{l})} = 
 \qquad \qquad \qquad \qquad   \qquad \qquad \qquad  \nonumber \\
\exp\left(-\ui \epsilon \sum_{s=1}^t \xi_{k_s l_s}\right)
\bra{\mathbf{k}}\left(\U^\star_\ua\right)^{\otimes t}\ket{\eta_1(\mathbf{k})}
\bra{\mathbf{l}}\left(\U^\star_\ub\right)^{\otimes t}\ket{\eta_1(\mathbf{l})}. \hspace*{0.6cm}
\end{align}
Hence for each term in the fourfold sum~\eqref{eq:spectral_form_factor_tensor_products}
the average over the phases and the two independent CUE($N$) matrices $\U_\ua$ 
and $\U_\ub$ factorizes.
We first evaluate the Haar average for subsystem $\ua$, while subsystem $\ub$ 
can be treated similarly.

More precisely, we aim for evaluating
\begin{align}
   f(\mathbf{i}, \mathbf{k})=\Big\langle \bra{\mathbf{i}}\U_\ua^{\otimes t}\ket{\eta_1(\mathbf{i})}
             \bra{\mathbf{k}}\left(\U^\star\right)_\ua^{\otimes t}\ket{\eta_1(\mathbf{k})} \Big\rangle.
   \label{eq:Haar_monomial}
\end{align}
Here, the expression to be averaged over the unitary group is a monomial in the 
matrix entries of $\U_\ua$ and $\U_\ua^\star$. 
The Haar average of such monomials (with the same number of matrix elements from $\U_\ua$ and $\U_\ua^\star$) can be expressed in terms of Weingarten functions $\Wg$ defined on the symmetric group $S_t$ \cite{ColSni2006}.
For a general monomial of the form $\bra{\mathbf{i}}\U_\ua^{\otimes t}\ket{\mathbf{j}}\bra{\mathbf{k}}\left(\U_\ua^\star\right)^{\otimes t}\ket{\mathbf{l}}$ the Haar average is non-zero only if $\mathbf{k}$ is a permutation $\mu$ of $\mathbf{i}$ and $\mathbf{l}$ is a permutation $\nu$ of $\mathbf{j}$.
The average is then given by \cite{Col2003,ColSni2006}
\begin{align}
    \Big\langle \bra{\mathbf{i}}\U_\ua^{\otimes t}\ket{\mathbf{j}}\bra{\mathbf{k}}\left(\U_\ua^\star\right)^{\otimes t}\ket{\mathbf{l}} \Big\rangle = \!\!
    \sum_{\mu,\nu\in S_t} \!\!
    \delta_{\mathbf{i},\mu(\mathbf{k})}\delta_{\mathbf{j},\nu(\mathbf{l})}
    \Wg (\nu\mu^{-1})
\end{align}
where the Kronecker $\delta$ is understood element wise.
Applied to the situation at hand, the Haar average of the monomial~\eqref{eq:Haar_monomial}
is non-zero only if there are permutations $\mu, \nu \in S_t$ such that
$\ket{\mathbf{i}} = \ket{\mu(\mathbf{k})}$ and $\ket{\eta_1(\mathbf{i})} = 
\ket{\nu\eta_1(\mathbf{k})}$, or equivalently 
$\ket{\mathbf{i}}=\ket{\eta_1^{-1}\nu\eta_1(\mathbf{k})}$.
That is, one has
\begin{align}
    f(\mathbf{i}, \mathbf{k})= \sum_{\mu,\nu\in S_t}
    \delta_{\mathbf{i},\mu(\mathbf{k})}\delta_{\mathbf{i},\eta_1^{-1}\nu\eta_1(\mathbf{k})}
    \Wg (\nu\mu^{-1}).
    \label{eq:Haar_average}
\end{align}
The Weingarten functions  $\Wg$ occurring in the above equations are 
rational functions of $N$ which decay as $N^{-t}$ if $\pi=\text{id}$ and at 
least as $N^{-(t+1)}$ otherwise as $N \to \infty$ \cite{Wei1978,Col2003,ColSni2006}.
This property allows for obtaining the leading contribution to the Haar average in the following.

As we are interested in this $N\to \infty$ limit we might assume $t\ll N$.
Consequently, the overwhelming majority of product states $\ket{\mathbf{i}}$ is such that the factors are pairwise distinct, $i_s \neq i_r$ for $s \neq r$.
In fact, the number of such states is given by $N!/((N-t)!)$ out of the total number $N^t$  of product basis states. 
Asymptotically these are all states as $N!/((N-t)!) \sim N^t$ for $N\to \infty$.
For such states we can relate the permutations $\mu$ and $\nu$ by
$\nu = \eta_1\mu\eta_1^{-1}$ and we can write $\Wg(\nu\mu^{-1})= 
\Wg(\eta_1\mu\eta_1^{-1}\mu^{-1})$ in Eq.~\eqref{eq:Haar_average}.
We therefore obtain
\begin{align}
f(\mathbf{i}, \mathbf{k})= \sum_{\mu \in 
S_t}\delta_{\mathbf{i},\mu(\mathbf{k})} 
\Wg(\eta_1\mu\eta_1^{-1}\mu^{-1}).
\end{align}
The leading terms correspond to permutations $\mu$ for which $\eta_1\mu\eta_1^{-1}\mu^{-1} = \text{id}$.
These are exactly the $t$ possible periodic shifts $\eta_r$ introduced above.
The asymptotics $\Wg(\text{id})\sim N^{-t} + 
\mathcal{O}\left(N^{-(t+1)}\right)$ then implies 
\begin{align}
f(\mathbf{i}, \mathbf{k})= N^{-t}\sum_{r = 0}^{t-1}\delta_{\mathbf{i},\eta_r(\mathbf{k})} + \sum_{\mu \in S_t}\delta_{\mathbf{i},\mu(\mathbf{k})}\mathcal{O}\left(N^{-(t+1)}\right).
\label{eq:haar_average_local}
\end{align}
An analogous expression holds for the Haar average over subsystem $\ub$.

Inserting those expressions for subsystem $\ua$ and $\ub$ into 
Eq.~\eqref{eq:spectral_form_factor_tensor_products} and keeping only the 
leading terms in $1/N$ we obtain
\begin{align}
    K(t) = N^{-2t}\sum_{\mathbf{i},\mathbf{j}}\sum_{r,s=0}^{t-1} \Big\langle \ue^{\ui \epsilon \mathbf{\theta}(\mathbf{i}, \mathbf{j}, \eta_r, \eta_s)}\Big\rangle.
\end{align}
Here, the double sums over $\mathbf{i}$ and $\mathbf{j}$ run over those states 
with pairwise distinct factors, the brackets denote the remaining average over 
the phases $\xi_{ij}$ and we define
\begin{align}
    \mathbf{\theta}(\mathbf{i}, \mathbf{j}, \mu, \nu) = \sum_{s=1}^t \xi_{i_s j_s} - \xi_{i_{\mu^{-1}(s)}j_{\nu^{-1}(s)}}.
    \label{eq:phases_for_averaging}
\end{align}
The average over the phases does not depend on $\mathbf{i}$ and $\mathbf{j}$ and hence the sum over these states gives a factor which asymptotically scales as $N^{2t}$ and exactly cancels the prefactor $N^{-2t}$.
Evaluating the average over the phases yields
\begin{align}
\Big\langle \ue^{\ui \epsilon \mathbf{\theta}(\mathbf{i}, \mathbf{j}, \eta_r, \eta_s)}\Big\rangle = \delta_{r,s} + (1-\delta_{r,s})|\chi(\epsilon)|^{2t},
  \label{eq:phases_averaged}
\end{align}
where
\begin{align}
\chi(\epsilon) = \big\langle \ue^{\ui \epsilon \xi } \big\rangle_\xi
\end{align}
denotes the characteristic function of the distribution of the phases $\xi_{ij}$.
Note that, for $r=s$ the phase $\mathbf{\theta}(\mathbf{i}, \mathbf{j}, \eta_r, \eta_s)=0$ and for $r \neq s$ all the $\xi_{ij}$ and $\xi_{i_{\eta_r^{-1}(s)}j_{\eta_s^{-1}(s)}}$ are different and independent from each other. 
This argument can be phrased more general by observing that the average gives 
$|\chi(\epsilon)|^{2k}$ where $k$ is the number of fixed points of 
$\eta_r^{-1}\eta_s$.
In the present case, of course, $k$ is either zero ($r \neq s$) or $t$ ($r=s$) but for higher moments of the spectral form factor the above observation becomes crucial; see Sec.~\ref{sec:sff_moments}.

Combining the above results we finally obtain
\begin{align}
    K(t) = t^2|\chi(\epsilon)|^{2t} + t(1 - |\chi(\epsilon)|^{2t})
    \label{eq:spectral_form_factor_semiclassics}
\end{align}
as $N \to \infty$.
For finite $N$ Eq.~\eqref{eq:spectral_form_factor_semiclassics} describes the leading order in $1/N$ of the spectral form factor for short times $t\ll t_{\text{SH}}$.
As $t^2 = K_N(t)^2$ as well as $t=K_{N^2}(t)$ for $t<N$ we can interpret Eq.~\eqref{eq:spectral_form_factor_semiclassics} as a time dependent convex combination of the spectral form factor of the uncoupled system and the spectral form factor for CUE($N^2$).
One might hope, that such a convex combination describes the spectral form factor of the coupled bipartite system even for times larger than the regime in which the derivation above is valid.
In this case the natural extension of the above result yields
\begin{align}
K(t) = |\chi(\epsilon)|^{2t}K_N(t)^2 + \left(1 - |\chi(\epsilon)|^{2t}\right) 
K_{N^2}(t)
\label{eq:spectral_form_factor_semiclassics2}
\end{align}
and extrapolates Eq.~\eqref{eq:spectral_form_factor_semiclassics} to a time regime $t\geq t_{\text{SH}}$.
Here we explicitly include the plateaus of $K_N(t)$ and $K_{N^2}(t)$ after the respective Heisenberg times.
Interestingly, a similar representation of the spectral form factor around the Thouless time, $t\approx t_\text{Th}$, very recently was obtained in a different setting realizing a minimal model of quantum glasses \cite{BarWinBalSwiGal2023:p}.
In Fig.~\ref{fig:sff_rmte_N50} we compare Eq.~\eqref{eq:spectral_form_factor_semiclassics2} with the numerically obtained spectral form factor.
As we choose uniformly distributed phases there, the characteristic function of
 their distribution is given by $\chi(\epsilon) = \sinc(\epsilon)$.
We find good agreement between the asymptotic result (black lines) 
and numerical data for times up to $t = t_{\text{SH}}$ ($\tau=\tau_{\text{SH}}$).
Surprisingly, the extrapolation~\eqref{eq:spectral_form_factor_semiclassics2} provides a 
reasonable description also for larger times.
In particular it explains the cusp at $\tau=\tau_{\text{SH}}$ for intermediate coupling, as at this time the spectral form factor for CUE$(N)$ sharply transitions from the ramp to the plateau.
Both the quality of this description and the time regime for which it describes the numerical data increase for stronger coupling.
Given the accuracy with which Eq.~\eqref{eq:spectral_form_factor_semiclassics2} describes the spectral form factor also  for times $t \geq t_{\text{SH}}$,  we exploit the latter in 
order to analyze the properties of the spectral form factor for the RMTE in the 
following, even though we currently lack a more rigorous derivation.

\section{Universality of the spectral form factor}
\label{sec:universality}

\begin{figure}[]
    \centering
    \includegraphics[width=8.5cm]{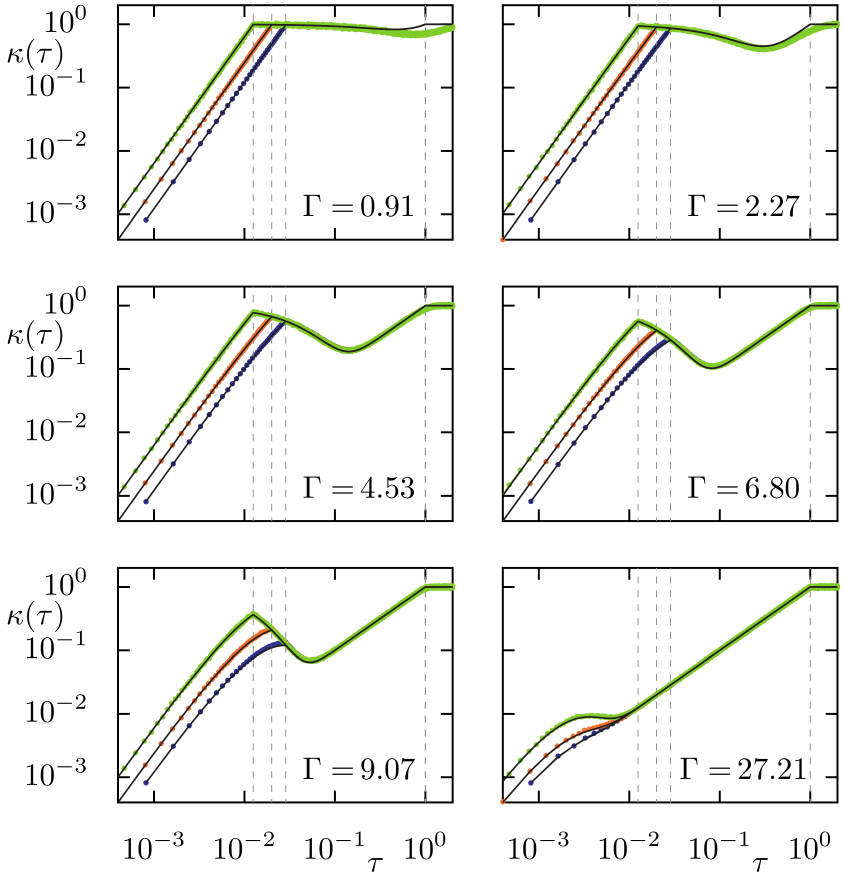}
    \caption{Rescaled spectral form factor $\kappa(\tau)$ for the RMTE with uniformly distributed phases for different $\Gamma$ (indicated in the individual panels) with $N=35$ (blue symbols, $200000$ realizations), $N=50$ (orange symbols, $20000$ realizations), and $N=80$ (green symbols, $6000$ realizations) in log-log scale with $N$ decreasing from top to bottom at small $\tau$.  The asymptotic result~\eqref{eq:spectral_form_factor_semiclassics2} is depicted as black lines. Dashed gray lines correspond to the Heisenberg times $\tau_{\text{SH}}$ of the subsystems and of the bipartite system $\tau_{\text{H}}=1$.}
    \label{fig:sff_rmte_universality}
\end{figure}

In chaotic quantum systems universality of spectral fluctuations typically refers to them being described by random matrix theory.
In case of the spectral form factor the corresponding random matrix description is provided by Eq.~\eqref{eq:spectral_form_factor_CUE}.
In the situation of coupled chaotic quantum systems with tunable coupling strength yet another notion of universality was introduced in Ref.~\cite{SriTomLakKetBae2016}.
There spectral fluctuations, and subsequently also entanglement properties of 
eigenstates \cite{LakSriKetBaeTom2016,TomLakSriBae2018,HerKieFriBae2020} as well as the entanglement production 
after a quench \cite{PulLakSriBaeTom2020,PulLakSriKieBaeTom2023} was found to depend on a universal 
scaling parameter $\Lambda(\epsilon, N)$ only.
The latter is a function of both coupling strength $\epsilon$ and the subsystems'
Hilbert space dimension $N$.
In close analogy we derive the universal dependence of the rescaled spectral form factor $\kappa(\tau)$ for times $t\geq t_{\text{SH}}$ ($\tau>\tau_{\text{SH}}$) on the scaling parameter 
\begin{align}
   \Gamma=\sigma \epsilon N, 
   \label{eq:scaling_parameter}
\end{align} 
i.e., the ratio of effective coupling strength $\sigma \epsilon$, with $\sigma^2$ the variance of the distribution of the phases $\xi_{ij}$ entering $\U_{\text{c}}$, and effective Planck's constant $1/N$.
In particular, this is independent of the concrete form of the distribution of phases.
This scaling parameter closely resembles the universal transition parameter obtained perturbatively in Ref.~\cite{SriTomLakKetBae2016}.
We comment on the subtle differences in Sec.~\ref{sec:perturbation_theory}, when 
adapting a perturbative treatment of the coupling.

In order to illustrate the above notion of universality we compute the spectral form factor numerically for the RMTE for various $\epsilon$ and $N$ but for fixed $\Gamma=\sigma\epsilon N$. 
This is depicted in Fig.~\ref{fig:sff_rmte_universality}, where we show numerical data for the RMTE with uniformly distributed phases $\xi_{ij}$ for six representative values of $\Gamma$ and three different system sizes respectively.
Indeed we observe, that for times $\tau>\tau_{\text{SH}}$ the rescaled spectral form factor 
depends only on the universal scaling parameter $\Gamma$.
In order to explain this universal behavior we rescale both time and spectral form factor in Eq.~\eqref{eq:spectral_form_factor_semiclassics2} by $N^2$ and obtain
\begin{align}
    \kappa(\tau) = |\chi(\epsilon)^{N^2}|^{2\tau} +  \left(1 - |\chi(\epsilon)^{N^2}|^{2\tau}\right)\kappa_{\text{CUE}(N^2)}(\tau) 
    \label{eq:sff_rescaled_semiclassics}
\end{align}
for $\tau \geq \tau_{\text{SH}}$.
Here $\kappa_{\text{CUE}(N^2)}(\tau) = \min\{\tau, 1\}$, and we used that the term $\kappa_{\text{CUE}(N)}^2(\tau)=1$ for this times.
By the central limit theorem applied to the characteristic function of the 
distribution of the phases one has
\begin{align}
    |\chi(\epsilon)|^{N^2}\approx \chi_{\mathcal{N}}(\sigma \epsilon N)
    \label{eq:central_limit_theorem}
\end{align}
for large $N$.
Here $\chi_{\mathcal{N}}(x) = \exp\left(-x^2/2\right)$
is the characteristic function of the standard normal distribution $\mathcal{N}$.
Accordingly, one has 
\begin{align}
    |\chi(\epsilon)|^{2t}=\exp\left(-\Gamma^2\tau\right).
    \label{eq:central_limit_theorem2}
\end{align}
Inserting this into Eq.~\eqref{eq:sff_rescaled_semiclassics} we find that $\kappa(\tau)$ depends only on the universal scaling parameter $\Gamma = \epsilon \sigma N$ implying the observed universality.
In particular this also holds in regimes where the spectral form factor does not follow the full random matrix result.
Hence it is specific to the bipartite setting considered here and therefore goes beyond the standard universal random matrix description of the composite system by CUE$(N^2)$.

\section{Thouless Time}
\label{sec:thouless_time}

For sufficiently large scaling parameter the rescaled spectral form 
factor will follow the linear ramp $\kappa(\tau)=\tau$ of the CUE($N^2$) 
spectral form factor after some initial time $t_{\text{Th}} <t_{\text{H}}$ 
($\tau_{\text{Th}}<\tau_{\text{H}}$). 
This time scale is referred to as Thouless time (in the context of many-body 
quantum systems) and its inverse sets the energy scale below which spectral 
correlations follow random matrix theory.

In order to derive the Thouless time from the spectral form factor we define $t_{\text{Th}}$ to be the time after which spectral form factor of the RMTE and CUE($N^2$) are close.
That is we require the difference of the spectral form factor from the random matrix result to be smaller than a given (small) threshold  $\Delta$. 
The latter might be measured in absolute units, $\Delta$ of order one, or alternatively in units of the Heisenberg time, $\Delta \sim  \delta N^2$ for $\delta \ll 1$.
We use both approaches in the following and begin with the latter.

In this case we define $t_{\text{Th}}$ to be 
 the solution of $K(t) - K_{N^2}(t) = K(t) - t = \delta N^2$  for 
small $\delta \ll 1$.
 For the asymptotic spectral form factor~\eqref{eq:spectral_form_factor_semiclassics2} this reads 
$\left(N^2 - t \right) |\chi(\epsilon)|^{2t} = \delta N^2$
We restrict to the case $t_{\text{SH}}<t_{\text{Th}}<t_{\text{H}}$ as on the 
one hand this is the regime, where the spectral form factor and hence rescaled 
Thouless time $\tau_{\text{Th}}$ depend on the scaling parameter $\Gamma$ only.
On the other hand, this is the regime, where one can solve for $\tau_{\text{Th}}$ exactly.
Using the universal dependence of the spectral form factor on $\Gamma$ and the central limit theorem for the characteristic function of the phases $\xi_{ij}$ yields the rescaled equation
\begin{align}
    \left(1 - \tau\right)\exp\left(-\Gamma^2 \tau\right)= \delta.
\end{align}
Its solution in the interval $\left(1/N, 1\right)$ can be expressed in terms of 
 the $0$-branch of the Lambert $\LambW$ function as
\begin{align}
\tau_{\text{Th}}=1 - \frac{1}{\Gamma^2}\LambW_0\!\left(\delta \Gamma^2 \ue^{\Gamma^2}\right).
\label{eq:thouless_time_analytic}
\end{align}

\begin{figure}[]
    \centering
    \includegraphics[width=8.5cm]{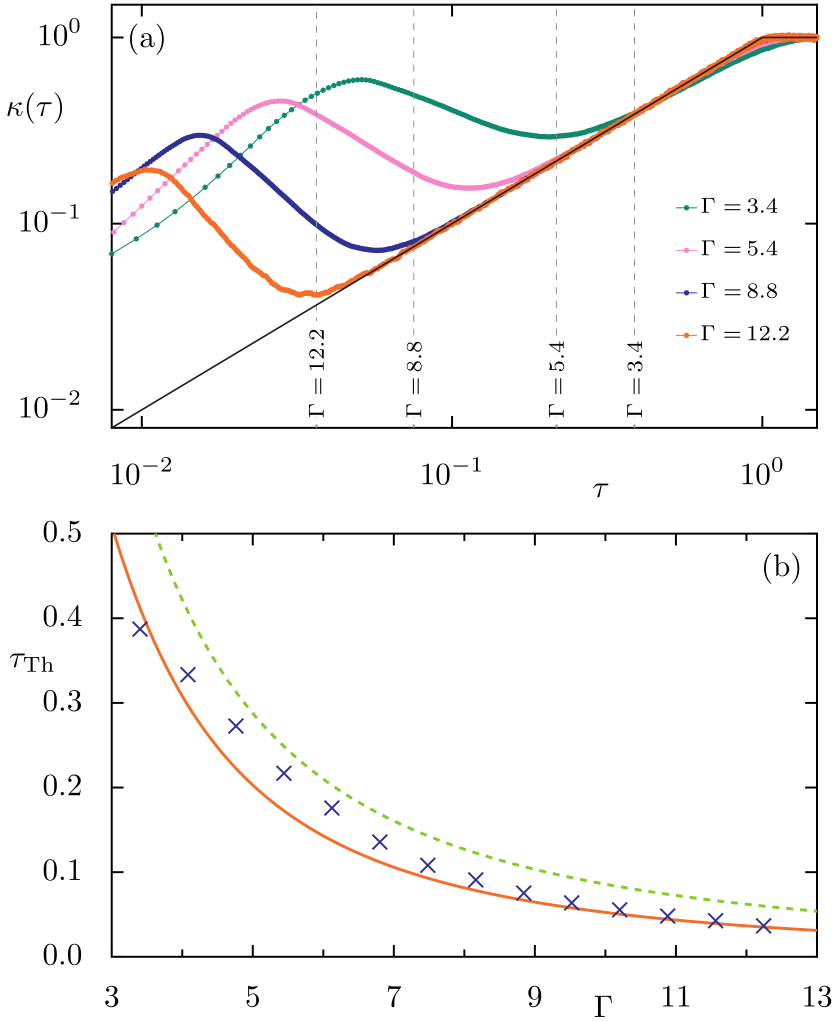}
    \caption{(a) Spectral form factor for different scaling parameter $\Gamma$ (decreasing from left to right). Connected symbols correspond to the numerically obtained, smoothed spectral form factor, while the solid black line depicts the random matrix result for CUE. The dashed lines indicate the numerically extracted Thouless times. (b) Thouless time $\tau_{\text{Th}}$ vs.~scaling parameter $\Gamma$ in the regime where $1/N < \tau_{\text{Th}} < 1$. Blue crosses correspond to numerical solutions $\kappa(\tau)-\tau<\delta$ while the solid orange line corresponds to Eq.~\eqref{eq:thouless_time_analytic} both for $\delta=0.005$. The dashed green line represents Eq.~\eqref{eq:Ehrenfest_time}.}
    \label{fig:rmte_thouless_time}
\end{figure}

We compare the above result with numerical solutions to $K(t) - t = \delta N^2$ in Fig.~\ref{fig:rmte_thouless_time}. 
In order to reduce the influence of the small scale fluctuations of the 
numerically obtained spectral form factor we smooth the numerical data by means 
of a moving time average over a small time window when extracting 
$t_{\text{Th}}$.
Although this introduces some artifacts at small $\tau$, the smoothing allows for a more reliable extraction of the much larger Thouless time $t_{\text{Th}}$.
We illustrate the smoothed spectral form factor in Fig.~\ref{fig:rmte_thouless_time}(a) and indicate the numerically obtained Thouless time for a few values of the scaling parameter $\Gamma$.
As demonstrated in Fig.~\ref{fig:rmte_thouless_time}(b) the numerically obtained rescaled Thouless time (blue crosses) is qualitatively well described by Eq.~\eqref{eq:thouless_time_analytic} (solid orange line) for large $\Gamma$.
For smaller $\Gamma$ the agreement is worse, which we attribute to the fact that for this regime the linear ramp is influenced by the deviations of the numerically obtained spectral form factor from its CUE($N^2$) counterpart around Heisenberg time $\tau = \tau_{\text{H}}$.
For larger $\Gamma$ than what is shown in Fig.~\ref{fig:rmte_thouless_time} the 
Thouless time is smaller than the Heisenberg time of the subsystems, 
$\tau_{\text{Th}}<\tau_{\text{SH}}$, and hence does not depend 
on $\Gamma$ only.
Moreover, there is no closed form expression for the Thouless time  for the asymptotic spectral form factor~\eqref{eq:spectral_form_factor_semiclassics} in this case.
We emphasize that the above properties are independent from the concrete choice of the arbitrary parameter $\delta$.

Let us also comment on the alternative definition of the Thouless time $t_{\text{Th}}$ as the solution to $K(t) - t = \Delta$, for $\Delta$ of order $1$,  instead.
Moreover consider again the regime  where $t_{\text{SH}} < t_{\text{Th}} \ll t_{\text{H}}$ , i.e., $N^2 - t \approx N^2$.
Solving for Thouless time yields
\begin{align}
t_{\text{Th}}=\frac{1}{2 \ln(|\chi(\epsilon)|)}\left(\ln(\Delta) - 2\ln(N)\right) \approx \frac{\ln(N)}{|\ln(|\chi(\epsilon)|)|}
\label{eq:Ehrenfest_time}
\end{align}
when $N$ is large.
This gives a rough qualitative approximation to the numerically obtained Thouless time, see Fig.~\ref{fig:rmte_thouless_time} (green dashed line).

In few particle systems with a well defined classical limit the Thouless time is often thought to be equal to the so-called Ehrenfest time $t_{\text{E}}$.
The latter is the time it takes for an initial, minimal uncertainty wave packet to spread over the whole system.
It can be estimated to scale with Planck's constant $1/N$ as $\ln(N)/K$ with $K$ the Kolmogorov-Sinai entropy of the underlying classical chaotic system given by the sum of positive Lyuapunov exponents.
Even though both time scales have similar scaling with $N$, they cannot be the same in the $\Gamma$ regime discussed above.
The simple reason for this claim is, that $t_{\text{Th}}>t_{\text{SH}}$, whereas the Ehrenfest time for chaotic subsystems has to be smaller than their respective Heisenberg times, i.e., $t_{\text{E}}<t_{\text{SH}}$.
Whether both time scales agree in the strong coupling regime or as $N \to \infty$, i.e., large $\Gamma$ when $t_{\text{Th}}<t_{\text{SH}}$ is, however, not ruled out by the above analysis.

\section{Fluctuations of the Spectral Form Factor}
\label{sec:sff_moments}

The spectral form factor is meaningful only upon averaging over an ensemble and hence one might study its statistical fluctuations in more detail.
In this section we provide an analysis of higher moments of the spectral form 
factor defined as
\begin{align}
    K_m(t) = \big\langle \big| \text{tr}\left(\U^t\right) \big |^{2m} \big\rangle  - N^{4m}\delta_{t0},
    \label{eq:moments_sff}
\end{align}
where, in particular $K(t)=K_1(t)$.
For the CUE($M$) the spectral form factor is exponentially distributed yielding \cite{ArgImrSmi1993,Pra1997,Kun1999}
\begin{align}
K_{Mm} = K_{\text{CUE}(M)m}(t) = m!K_ {M}(t)^m,
\end{align}
which motivates the definition of the rescaled moments 
\begin{align}
    \kappa_m(\tau) = \frac{1}{N^2}\left(\frac{K_m(t)}{m!} \right)^{1/m},
    \label{eq:moments_sff_rescaled}
\end{align}
where again $\tau=t/N^2$.

\begin{figure}[]
    \centering
    \includegraphics[width=8.5cm]{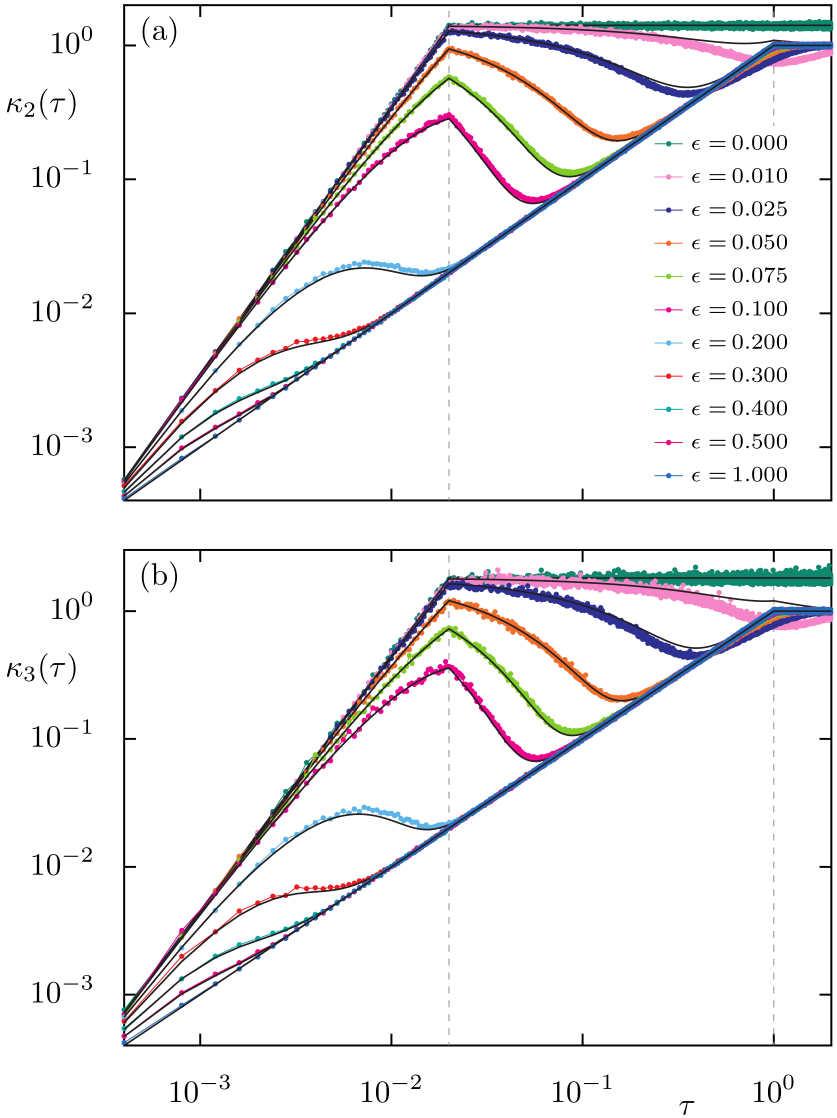}
    \caption{Rescaled second (a) and third (b) moment of the spectral form 
factor $\kappa(\tau)$ for the RMTE at $N=50$ and different coupling strengths 
$\epsilon$ (see legend, increasing from top to bottom) in log-log scale. Colored symbols correspond to 
numerical data obtained from $20000$ realizations of the RMTE with uniformly 
distributed phases $\xi_{ij}$. The asymptotic 
results~\eqref{eq:moments_semiclassics_3} and~\eqref{eq:moments_semiclassics_4} are depicted as black lines.
        Dashed gray lines correspond to the Heisenberg time $\tau_{\text{SH}}$ of the subsystems and of the bipartite system $\tau_{\text{H}}$.}
    \label{fig:sff_moment23_rmte_N50}
\end{figure}

In Fig.~\ref{fig:sff_moment23_rmte_N50} we show the second and third rescaled moment for the RMTE used in Fig.~\ref{fig:sff_rmte_N50}.
Due to the rescaling we find similar qualitative behavior of the moments as a function of the coupling strength $\epsilon$ as for the spectral form factor depicted in Fig.~\ref{fig:sff_rmte_N50}.
Note, however, that even after rescaling the moments do not match the spectral form factor exactly.
Most notably this is the case for the uncoupled system, for which $K_m(t)=K_ {\text{CUE}(M)m}(t)^2$ leading to an initial quadratic growth of $\kappa_m(\tau)$ and subsequent plateau $\kappa_m(\tau)=(m!)^{1/m}>1$ for $\tau\geq \tau_{\text{SH}}$.
For nonzero coupling one has saturation at large $\tau$ to the value $\kappa_m(\tau)=1$, matching the exponential distribution of the CUE spectral form factor.
Additionally an exponential distribution is observed also for the regime of the linear ramp for sufficiently strong coupling and times larger than the Thouless time.

In order to describe higher moments of the spectral form factor in the limit $N \to \infty$ we extend the methods from Sec.~\ref{sec:sff_semiclassics}.
The computation for $m>1$ is similar in spirit as for $m=1$ 
presented there but some details are more involved. 
The derivation resembles the computation of moments of the spectral form 
factors in a random quantum circuit setting with large local Hilbert space 
dimension \cite{ChaDeCha2021}.
In contrast to this many-body setting, which allows for asymptotic results in the thermodynamic limit of a large number of lattice site (subsystems), we obtain exact results in the bipartite setting.

The starting point is the generalization of Eq.~\eqref{eq:spectral_form_factor_tensor_products} for $K_m(t)$.
This is obtained by replacing all the $t$-fold tensor products by $mt$-fold tensor products
and by considering the representation of $S_{mt}$ which permutes tensor factors on the enlarged Hilbert space $\left(\mathcal{H}^{\otimes t}\right)^{\otimes m}$.
By taking the $m$-fold tensor product of the corresponding representation of 
$S_t$ on $\mathcal{H}^{\otimes t}$ we can represent the cyclic $t$-periodic 
shifts $\eta_r$ also on the larger Hilbert space and denote it by 
$\eta_r^{\otimes m}$ and interpret it as an element of $S_{mt}$. 
That is $\eta_r^{\otimes m}$ shifts the first $t$ factors (the first copy) periodically by $r$ as 
well as the second $t$ factors (the second copy), etc. 

This allows for rewriting the $m$-th moment as 
\begin{align}
K_m(t) & = \sum_{\mathbf{i},\mathbf{j},\mathbf{k},\mathbf{l}}\Big\langle
\bra{\mathbf{i},\mathbf{j}}\U^{\otimes mt}\ket{\eta_1^{\otimes m}(\mathbf{i}),\eta_1^{\otimes m}(\mathbf{j})} 
 \nonumber \\
 & \qquad \quad \times \bra{\mathbf{k},\mathbf{l}}\left(\U^\star \right)^{\otimes mt}\ket{\eta_1^{\otimes m}(\mathbf{k}),\eta_1^{\otimes m}(\mathbf{l})} \Big\rangle,
\label{eq:spectral_form_factor_moment_tensor_products}
\end{align}
where the sums run over the product basis in $\mathcal{H}^{\otimes mt}$.
From here one might proceed in an analogous way as for the case $m=1$.
Again the averages over the two independent CUE($N$) and the phases of the coupling factorize.
The leading contribution for the CUE($N$) average is again determined by the 
asymptotics of the Weingarten function.
The condition for a permutation $\mu \in S_{mt}$ to give rise to a term which does not vanish as $N \to \infty$ becomes $\mu = \eta_1^{\otimes m}\mu \left(\eta_1^{-1}\right)^{\otimes m}$.
A simple set of solutions is given by permutations which implement $t$-periodic shifts independently in each of the $m$ copies.
Another simple set of solutions is given by permutations of the individual copies.
In fact all the solutions to the constraint are given by combinations of the above simple solutions as we show in App.~\ref{App:Permutations}.
More formally, we demonstrate there that solutions form a subgroup of $S_{mt}$ isomorphic to the semidirect product $G_m = S_m \ltimes \langle \eta_1 \rangle^m$.
Here the second factor denotes the $m$-fold direct product of the cyclic group generated by $\eta_1$ and  $S_m$ acts on that product group by permuting the factors in the $m$-fold product. We identify $G_m$ with the corresponding subgroup of $S_{mt}$.
Keeping only the leading terms for the two CUE($N$) averages yields
\begin{align}
    K_m(t) = N^{-2mt}\sum_{\mathbf{i},\mathbf{j}}\sum_{\mu,\nu\in G_m}\Big\langle \ue^{\ui \epsilon \mathbf{\theta}(\mathbf{i},\mathbf{j}, \mu, \nu)} \Big\rangle.
\end{align}
Here the phase $\mathbf{\theta}(\mathbf{i},\mathbf{j}, \mu, \nu)$ is given by 
Eq.~\eqref{eq:phases_for_averaging} with the sum running up to $mt$ instead of just $t$.
The remaining average over the phases is again independent from the states $\mathbf{i}$ and $\mathbf{j}$ and the sum over the states gives an factor which cancels the prefactor $N^{-2mt}$ as $N \to \infty$.

The same argument leading to Eq.~\eqref{eq:phases_averaged} shows that
the average over the phases depends only on the number of fixed points of
$\mu \nu^{-1}$, which is of the form $kt$ with $k\in \{0,\ldots,m\}$.
The average is consequently given by 
\begin{align}
    \Big\langle \ue^{\ui \epsilon \mathbf{\theta}(\mathbf{i},\mathbf{j}, \mu, \nu)} \Big\rangle = |\chi(\epsilon)|^{2t(m - k)}.
    \label{eq:phase_average}
\end{align}
We therefore can rewrite the $m$-th moment as
\begin{align}
K_m(t)  = m!t^m\sum_{k=0}^{m}A_k(t)|\chi(\epsilon)|^{2t(m - k)},
\label{eq:moments_semiclassics_1}
\end{align}
where $A_k(t)$ denotes the number of permutations in $G_m$ with exactly $kt$ fixed points.
The prefactor originates from a change of the summation over $\mu$ to $\mu^\prime = \mu\nu^{-1}$ after which the sum over $\nu$ becomes trivial.
The $A_k(t)$ are computed in App.~\ref{App:FixedPoints} and are given by the sum
\begin{align}
    A_k(t) = \sum_{l=k}^{m}\binom{m}{l}\binom{l}{k}!(m-l)t^{m-l}(t-1)^{l-k},
    \label{eq:A_kt}
\end{align}
which is a polynomial in $t$ of degree at most $m$.
Here $!x$ denotes the subfactorial.
Although Eq.~\eqref{eq:moments_semiclassics_1} is not in a closed form it exactly reproduces the random matrix result for the uncoupled case as $\sum_kA_k=|G_m|=m!t^m$.
Moreover, at non-zero coupling with $|\chi(\epsilon)|<1$ and for large times all but the term corresponding to $k=m$ are exponentially suppressed. 
As $A_m = 1$ we hence recover the random matrix result for CUE($N$) at large time, i.e., we
obtain $K(t) = m!t^m$.
That is, we expect the spectral form factor to be exponentially distributed at sufficiently large times $t\gtrsim t_{\text{Th}}$.

A straight forward calculation shows that Eq.\eqref{eq:moments_semiclassics_1} reproduces Eq.~\eqref{eq:spectral_form_factor_semiclassics} for $m=1$.
For $m=2$ we obtain
\begin{align}
K_2(t) = & 2t^2\left(1 - 2|\chi(\epsilon)|^{2t} + |\chi(\epsilon)|^{4t}\right)
        + \left(2t^2\right)^2|\chi(\epsilon)|^{4t} \nonumber \\
&   + 2t^3 \left(2|\chi(\epsilon)|^{2t} - 2|\chi(\epsilon)|^{4t}\right).
\label{eq:moments_semiclassics_2}
\end{align}
This gives a good description of the rescaled second moment for times $t<t_{\text{SH}}$ as it 
is depicted in Fig.~\ref{fig:sff_moment23_rmte_N50}(a).

Moreover, we observe, that Eq.~\eqref{eq:moments_semiclassics_2} is a time dependent convex combination of $2t^2=K_{N^22}(t)$, $4t^4 = K_{N2}(t)^2$ and $2t^3$.
Again one might assume, that such a convex combination describes the second moment also for times $t>t_{\text{SH}}$ as it was demonstrated for the first moment.
In general this requires to replace monomials $t^l$ by products of moments of the spectral form factor for $\text{CUE}(N^2)$ and $\text{CUE}(N)$, where the latter should enter to even powers given the bipartite setting.
Given the exponential distribution of the CUE spectral form factor we can restrict ourselves to products of $K_{N^2}(t)$ and $K_N(t)^2$.
For the first two monomials the obvious choice is $2t^2=2K_{N^2}^2(t)$, and $4t^4 = 4K_N(t)^4$ as indicated above.
For the third monomial $2t^3$ an analogous substitution is not obvious.
We choose the replacement $2t^3 = 2K_{N^2}(t)K_N(t)^2$. 
We argue below, that this is the only choice consistent with observations from numerical data.
With the above substitutions the second moment is given by 
\begin{align}
K_2(t) =& \,  2K_{N^2}(t)^2 \left(1 - |\chi(\epsilon)|^{2t}\right)^2 + 4 K_N(t)^4 |\chi(\epsilon)|^{4t} \nonumber \\
& + 4K_{N^2}(t)K_N(t)^2|\chi(\epsilon)|^{2t}\left(1 - |\chi(\epsilon)|^{2t}\right). 
\label{eq:moments_semiclassics_3}
\end{align}
We compare this extrapolated result with the numerically obtained data in Fig.~\ref{fig:sff_moment23_rmte_N50}(a).
The agreement for the rescaled second moment $\kappa_2(\tau)$ is quantitatively 
slightly worse but qualitatively essentially similar to what we observe for the 
rescaled spectral form factor,
see Fig.~\ref{fig:sff_rmte_N50} and the corresponding discussion.

In complete analogy for the third moment we obtain
\begin{align}
K_3(t) =& \,  6K_{N^2}(t)^3 \left(1 - |\chi(\epsilon)|^{2t}\right)^3 + 36K_N(t)^6 |\chi(\epsilon)|^{6t} \nonumber \\
& + 18K_{N^2}(t)^2K_N(t)^2|\chi(\epsilon)|^{2t}\left(1 - |\chi(\epsilon)|^{2t}\right)^2 \nonumber \\
& + 36K_{N^2}(t)K_N(t)^4|\chi(\epsilon)|^{4t}\left(1 - |\chi(\epsilon)|^{2t}\right),
\label{eq:moments_semiclassics_4}
\end{align}
where the first terms correspond to the third moment of the spectral form factor of $\text{CUE}(N^2)$ and the second term to the square of the third moment for $\text{CUE}(N)$.
In Fig.~\ref{fig:sff_moment23_kicked_rotor_N50}(b) we compare this with the numerically 
obtained third moment and again find similar agreement as for the first and second moment.
Similar agreement is obtained for higher moments as well (not shown).
So far, Eqs.~\eqref{eq:moments_semiclassics_3}~and~\eqref{eq:moments_semiclassics_4} are justified only by their agreement with numerical data.
As it is the case for the first moment, a thorough derivation is beyond the scope of the methods used to derive the leading contribution for finite $N$ and small times $t < t_{\text{SH}}$.

\begin{figure}[]
    \centering
    \includegraphics[width=8.5cm]{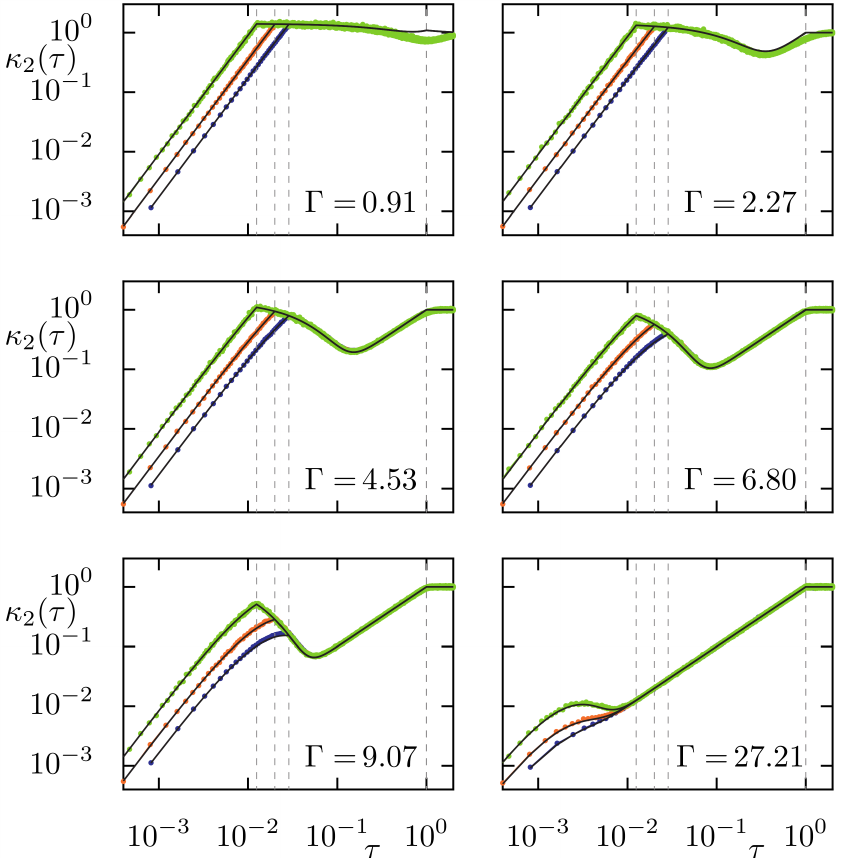}
    \caption{Rescaled second moment of the spectral form factor $\kappa_2(\tau)$ for the RMTE with uniformly distributed phases for different $\Gamma$ (indicated in the individual panels) with $N=35$ (blue symbols, $200000$ realizations), $N=50$ (orange symbols, $20000$ realizations), and $N=80$ (green symbols, $6000$ realizations) in log-log scale with $N$ decreasing from top to bottom at small $\tau$.  The asymptotic result~\eqref{eq:moments_semiclassics_3} is depicted as black lines. Dashed gray lines correspond to the Heisenberg times $\tau_{\text{SH}}$ of the subsystems and of the bipartite system $\tau_{\text{H}}$.}
    \label{fig:sff_moment2_rmte_universality}
\end{figure}

\begin{figure}[]
    \centering
    \includegraphics[width=8.5cm]{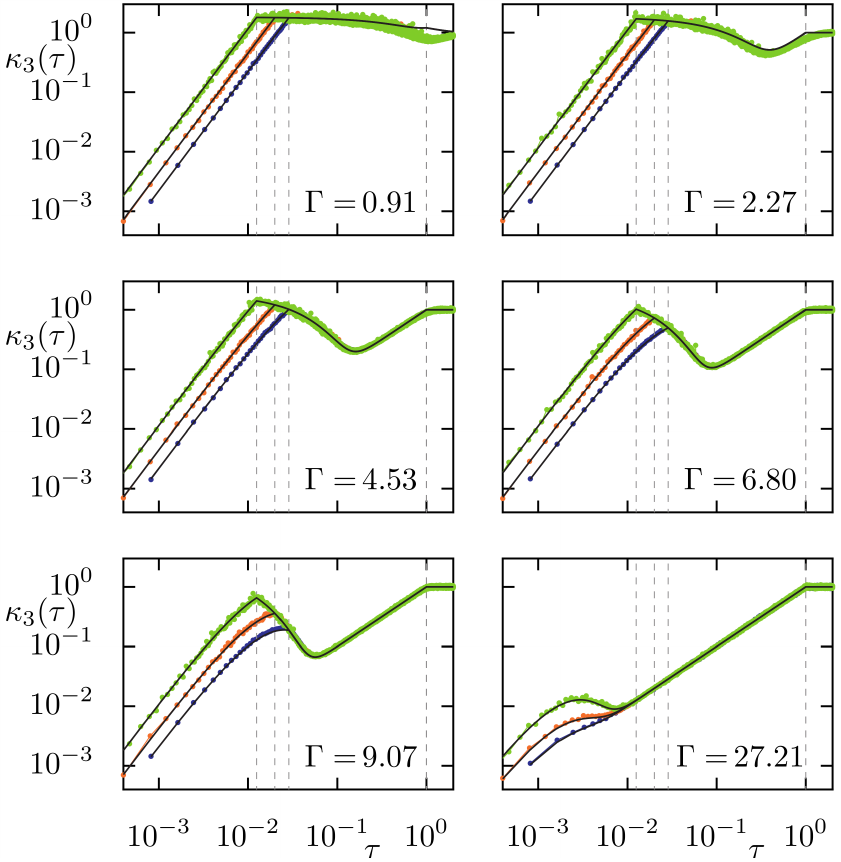}
    \caption{Same as Fig.~\ref{fig:sff_moment2_rmte_universality} but for the third moment of the spectral form factor. The asymptotic result (black lines) is given by Eq.~\eqref{eq:moments_semiclassics_4}.}
    \label{fig:sff_moment3_rmte_universality}
\end{figure}

Assuming the conjectured form of the moments of the spectral form factor above, we find universal dependence of the rescaled moments on the scaling parameter $\Gamma$ for times $\tau > \tau_{\text{SH}}$.
Again this is a consequence of the central limit theorem applied to the characteristic function $\chi(\epsilon)$ as in the $m=1$ case.
Universality  in the sense of Sec.~\ref{sec:universality} for the RMTE is confirmed in Figs.~\ref{fig:sff_moment2_rmte_universality}~and~\ref{fig:sff_moment3_rmte_universality}.
There we show the second and third rescaled moment, respectively, for different $N$ while keeping $\Gamma$ fixed.
Both the numerical data and the corresponding asymptotical results collapse on 
a single curve for $\tau>\tau_{\text{SH}}$.

The above explanation for universality might be reversed in order to fix the substitution $t^l=K_{N^2}(t)^ kK_N(t)^{l - k}$ by requiring that the extrapolated result gives rise to the numerically observed universal dependence on $\Gamma$.
Here only monomials $t^l$ with $m \leq l \leq 2m$ appear in Eq.~\eqref{eq:moments_semiclassics_1} and the bipartite setting requires $l - k$ to be even.
In order for the rescaled moment to depend on $\tau=t/N^2$ and $\Gamma$ only one needs $K_{N^2}(t)^ kK_N(t)^{l - k}/N^{2m} = \tau^k$ for $\tau_{\text{SH}} < \tau < \tau_{\text{H}}$.
The only consistent choice is $k=2m - l$, which indeed is used in Eqs.~\eqref{eq:moments_semiclassics_3}~and~\eqref{eq:moments_semiclassics_4}.
Therefore at least the substitutions required to extrapolate Eq.~\eqref{eq:moments_semiclassics_1} and its analog for higher moments is uniquely fixed by the observed universality.
To summarize, also higher moments of the spectral form factor are given by a time dependent convex combination of the uncoupled case, the strongly coupled case, and terms involving products of lower moments which at times $\tau > \tau_{\text{SH}}$ depends on $\Gamma$ only.

\section{Spectral Form Factor for Coupled Kicked Rotors}
\label{sec:kicked_rotors}

So far we considered only the RMTE tailored for discussing spectral fluctuation 
for coupled chaotic quantum systems with tunable coupling strength.
In this section, in contrast, we demonstrate that the theory derived above is capable of describing spectral correlations also for more realistic settings.
In particular we will use a system of coupled kicked rotors \cite{Chi1979,Fro1972} to validate the predictions from the RMTE.
For the case of the level spacing distribution, the applicability of the RMTE model and universality is well established \cite{SriTomLakKetBae2016}.
We extend this to the spectral form factor using the large $N$ asymptotics obtained in Sec.~\ref{sec:sff_semiclassics}.

The coupled quantum kicked rotors have been studied extensively in various 
contexts  
\cite{AdaTodIke1988,Lak2001,RicLanBaeKet2014,SriTomLakKetBae2016,
LakSriKetBaeTom2016,TomLakSriBae2018} 
and have been realized experimentally \cite{GadReeKriSch2013}.
They are the quantization of the corresponding two 
degree of freedom classical dynamical system with toric phase space represented 
by $[0, 1]^4$ with periodic boundary conditions. 
The time dependent, i.e., periodically kicked Hamiltonian of the coupled kicked rotors reads \cite{Fro1972}
\begin{align}
    H(t) = \frac{p_\ua^2}{2} + \frac{p_\ub^2}{2} + V(q_\ua, q_\ub)\!\sum_{n 
=-\infty}^\infty \delta(t - n).
\end{align}
Here $V(q_\ua, q_\ub)=V_\ua(q_\ua) + V_\ub(q_\ub) +  V_{\text{c}}(q_\ua, 
q_\ub)$ is the potential energy given by the single particle potentials
\begin{align}
V_i(q_i) = \frac{k_i}{4\pi^2}\cos(2\pi q_i)
\end{align}
for $i \in \{\ua, \ub\}$ and the coupling potential
\begin{align}
V_{\text{c}}(q_\ua, q_\ub) = \frac{\gamma}{4\pi^2}\cos(2\pi [q_\ua + q_\ub]),  
\end{align}
whose strength is given by the parameter $\gamma$.
We fix the kick strength of the subsystems as $k_\ua=9.7$ and $k_\ub=10.5$ for 
which classical dynamics is fully chaotic with possible regular islands being 
negligible small.

\begin{figure}[]
    \centering
    \includegraphics[width=8.5cm]{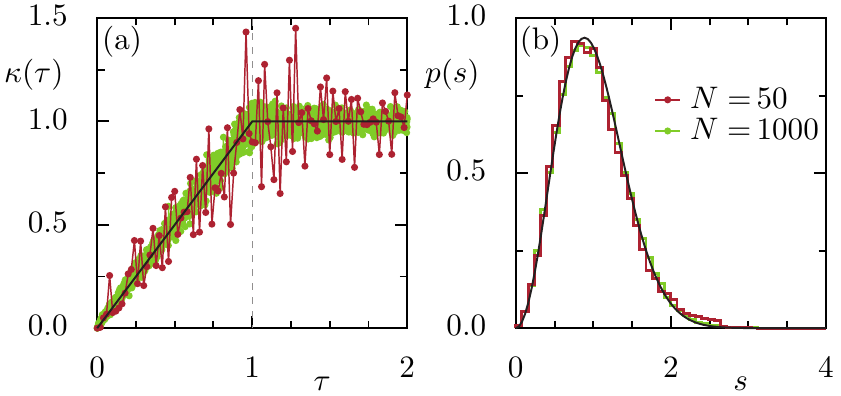}
    \caption{(a) Rescaled spectral form factor $\kappa(\tau)=K(t)/N$, $\tau=t/N$, for a single kicked rotor with $k=9.7$ at $N=50,1000$ ($10^6, 10^4$ realizations) in linear scale.
         The random matrix result~\eqref{eq:spectral_form_factor_CUE} is depicted as a black line. The dashed gray line correspond to the Heisenberg time $\tau_{\text{H}}=1$.
         (b) Level spacing distribution corresponding to (a) for $10^5$ ($N=50$) and $10^2$ ($N=1000$) realizations. The black line corresponds to Eq.~\eqref{eq:wigner_surmise}.
         Results for $k=10.5$ are not depicted as they are similar to $K=9.7$.}
    \label{fig:sff_single_kicked_rotor}
\end{figure}

Quantization of the classical system requires the effective Planck's constant to be of the form $h=1/N$ for integer $N$.
Time evolution is given by a
Floquet operator of the form~\eqref{eq:rmte} acting on the finite dimensional
Hilbert space $\mathcal{H} =\mathcal{H}_\ua \otimes \mathcal{H}_\ub = 
\mathds{C}^N \otimes \mathds{C}^N \simeq \mathds{C}^{N^2}$ of dimension $\dim 
\mathcal{H}=N^2$.
The time evolution within the individual subsystems is governed by \cite{BerBalTabVor1979,HanBer1980,ChaShi1986,KeaMezRob1999,DegGra2003}
\begin{align}
    \U_i = \ue^{- \pi \ui N p_i^2} \ue^{-2 \pi \ui N V_i(q_i)}
\end{align}
for $i \in \{\ua, \ub \}$ whereas the coupling reads
\begin{align}
\U_{\text{c}}= \ue^{-2 \pi N \ui V_{\text{c}}(q_\ua, q_\ub)}  = 
\ue^{-\frac{\ui \gamma N}{2\pi}\cos(2\pi[q_\ua + q_\ub])}.
\end{align}
For each subsystem $i \in \{\ua, \ub \}$ the Hilbert space $\mathcal{H}_i$ is 
spanned by either the position eigenstates $\ket{q_{i,n}}$ with eigenvalues  \cite{KeaMezRob1999,DegGra2003}
\begin{align}
    q_{i, n} = \frac{1}{N}(n + \theta_i^q) \in [0, 1)
    \end{align}
or by the momentum eigenstates $\ket{p_{i,n}}$ with eigenvalues
\begin{align}
    p_{i, n} = \frac{1}{N}(n + \theta_i^p) \in [0, 1)
\end{align}
for $n \in \{0,\ldots, N-1\}$, respectively.
The total Hilbert space $\mathcal{H}$ of the bipartite system is spanned by the corresponding 
product basis. 
The vector $\mathbf{\theta} = \left(\theta_\ua^q,\theta_\ub^q, 
\theta_\ua^p,\theta_\ub^p  \right)
\in [0, 1)^4$  of Bloch phases determines the boundary conditions for quantum states.
For phases $\theta_i^q,\theta_i^p \notin \{0, 1/2\}$ the corresponding subsystem exhibits no anti-unitary symmetry, e.g., time-reversal symmetry, and hence falls in the unitary symmetry class.
Note, that on the one hand for this choice of phases the evolution operator of the coupled kicked rotors does not provide a proper quantization of the classical kicked rotor on the torus \cite{KeaMezRob1999}.
On the other hand it provides the most convenient ensemble for averaging (see below) and was used also in earlier studies \cite{SriTomLakKetBae2016,LakSriKetBaeTom2016,TomLakSriBae2018}.
We confirm, that spectral fluctuations for the individual kicked rotors are indeed  well described by the CUE($N$) by comparing their spectral form factor with the random matrix result Eq.~\eqref{eq:spectral_form_factor_CUE}.
To this end we choose the Bloch phases as i.i.d. random variables uniformly distributed in $[0, 1)$ and average over the resulting ensemble of kicked rotors.
The resulting spectral form factor is depicted in
Fig.~\ref{fig:sff_single_kicked_rotor}(a) (colored connected symbols) and follows the random matrix result for all times.
This is even the case for relatively small $N=50$ which is used below for the coupled system.
However, for $N=50$ fluctuations around the random matrix result are much larger than for $N=1000$.
Additionally we depict the corresponding level spacing distributions $p(s)$ in 
Fig.~\ref{fig:sff_single_kicked_rotor}(b), which also follow the random matrix result~\eqref{eq:wigner_surmise}.

\begin{figure}[]
    \centering
    \includegraphics[width=8.5cm]{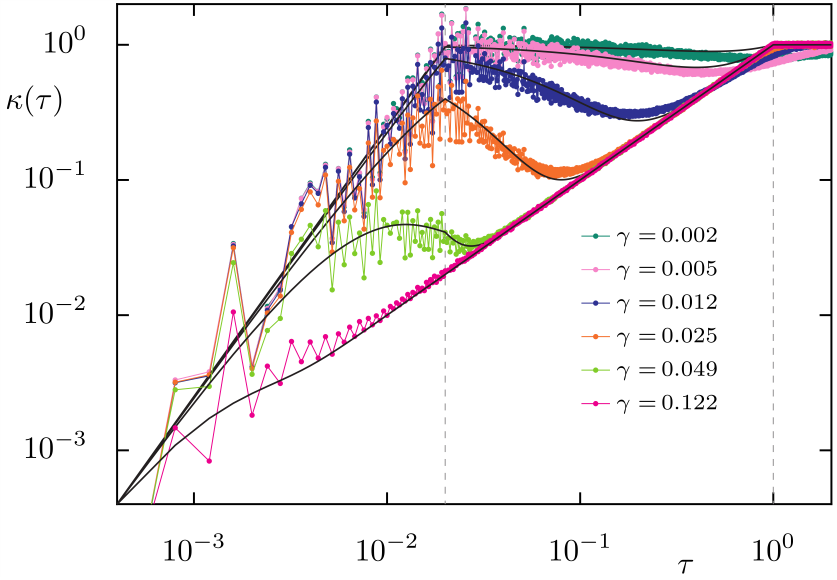}
    \caption{Rescaled spectral form factor $\kappa(\tau)$ for the coupled 
kicked rotors at $N=50$ and different coupling strengths $\gamma$ (see legend, increasing from top to bottom) 
in log-log scale. Colored symbols correspond to numerical data obtained from 
$10000$ realizations averaged over the Bloch phases. The asymptotic 
result~\eqref{eq:spectral_form_factor_semiclassics2} is depicted as black lines.
       Dashed gray lines correspond to the Heisenberg time $\tau_{\text{SH}}$ of the subsystems and of the bipartite system $\tau_{\text{H}}$.}
    \label{fig:sff_kicked_rotor_N50}
\end{figure}

\begin{figure}[]
    \centering
    \includegraphics[width=8.5cm]{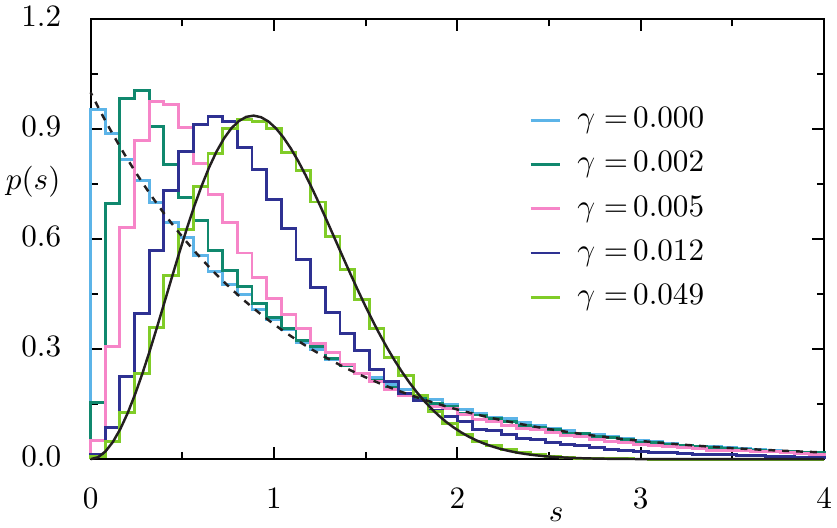}
    \caption{Level spacing distribution for the couple kicked rotor at $N=50$ 
and different coupling strengths $\gamma$ (see legend, increasing from left to right). Colored histograms correspond to numerical data obtained from 100 realizations with i.i.d. 
uniformly distributed Bloch phases. The solid black line represents the random 
matrix result for the CUE, Eq.~\eqref{eq:wigner_surmise}, whereas the dashed black line represents Poissonian 
statistics.}
    \label{fig:level_spacing_kicked_rotor_N50}
\end{figure}

\begin{figure}[]
    \centering
    \includegraphics[width=8.5cm]{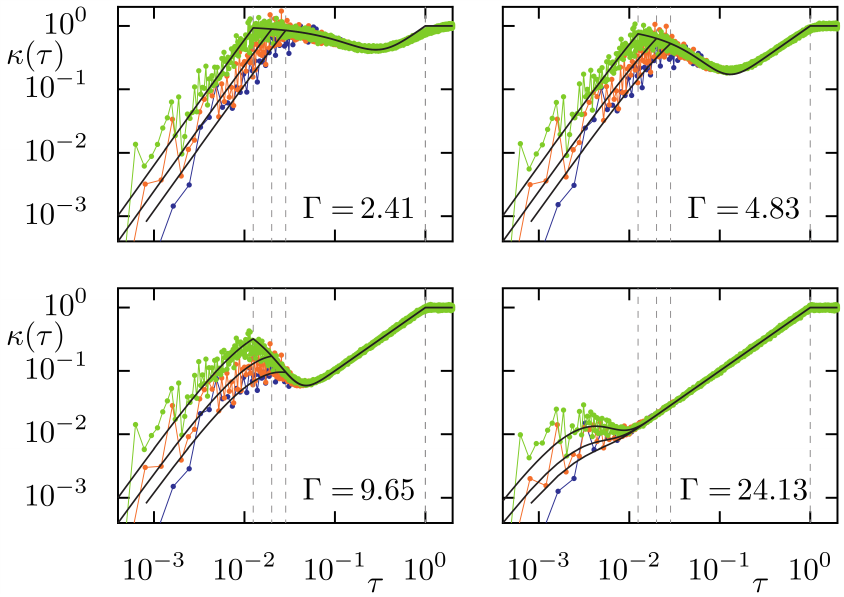}
    \caption{Rescaled spectral form factor $\kappa(\tau)$ for the coupled kick rotors for different $\Gamma$ (indicated in the individual panels) with $N=35$ (blue symbols, $10000$ realizations), $N=50$ (orange symbols, $10000$ realizations), and $N=80$ (green symbols, $1000$ realizations) in log-log scale with $N$ decreasing from top to bottom at small $\tau$.  The asymptotic result~\eqref{eq:spectral_form_factor_semiclassics2} is depicted as black lines. Dashed gray lines correspond to the Heisenberg times $\tau_{\text{SH}}$ of the subsystems and of the bipartite system $\tau_{\text{H}}$.}
    \label{fig:sff_kicked_rotor_universality}
\end{figure}

In order to compute the spectral form factor numerically for the coupled system we  again define an ensemble of coupled kicked rotors at fixed coupling strength $\gamma$ by taking the four Bloch phases to be i.i.d. random variables uniformly distributed in $[0, 1)$.
We depict the resulting spectral form factor for $N=50$ and various coupling strengths $\gamma$ in Fig.~\ref{fig:sff_kicked_rotor_N50}.
Qualitatively the spectral form factor shows similar behavior as for the RMTE.
Notable differences, however, are the period two oscillations on top of the 
average behavior as well as stronger fluctuations.
The latter roughly correspond to the size of fluctuations of the spectral form factor for a single kicked rotor at $N=50$, see Fig.~\ref{fig:sff_single_kicked_rotor}(a).
As in the case of the RMTE we contrast the transition of the spectral form factor with that of the level spacing distribution in Fig.~\ref{fig:level_spacing_kicked_rotor_N50}.
We find a similar transition from the exponential distribution of the uncoupled kicked rotors towards Eq.~\eqref{eq:wigner_surmise}, which is again completed for smaller coupling strenght $\gamma = 0.049$ than the complete transition of the spectral form factor towards the CUE$(N^2)$ result~\eqref{eq:spectral_form_factor_CUE}.

\begin{figure}[]
    \centering
    \includegraphics[width=8.5cm]{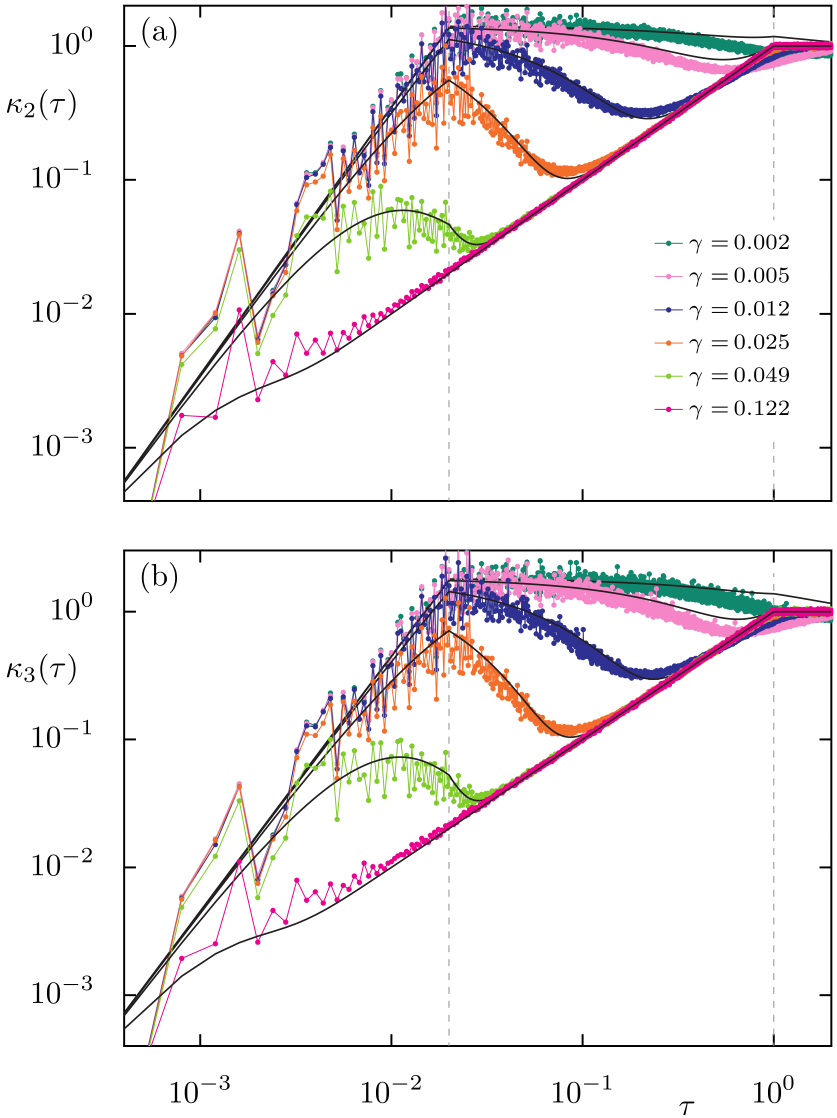}
    \caption{Rescaled second (a) and third (b) moment of the spectral form 
factor $\kappa(\tau)$ for the coupled kicked rotors at $N=50$ and different 
coupling strenghts $\gamma$ (see legend, increasing from top to bottom) in log-log scale. Colored symbols 
correspond to numerical data obtained for the same ensemble as in Fig.~\ref{fig:sff_kicked_rotor_N50}. The asymptotic 
        results~\eqref{eq:moments_semiclassics_3} and~\eqref{eq:moments_semiclassics_4} are depicted as black lines.
        Dashed gray lines correspond to the Heisenberg time $\tau_{\text{SH}}$ of the subsystems and of the bipartite system, $\tau_{\text{H}}$.}    \label{fig:sff_moment23_kicked_rotor_N50}
\end{figure}

\begin{figure}[]
    \centering
    \includegraphics[width=8.5cm]{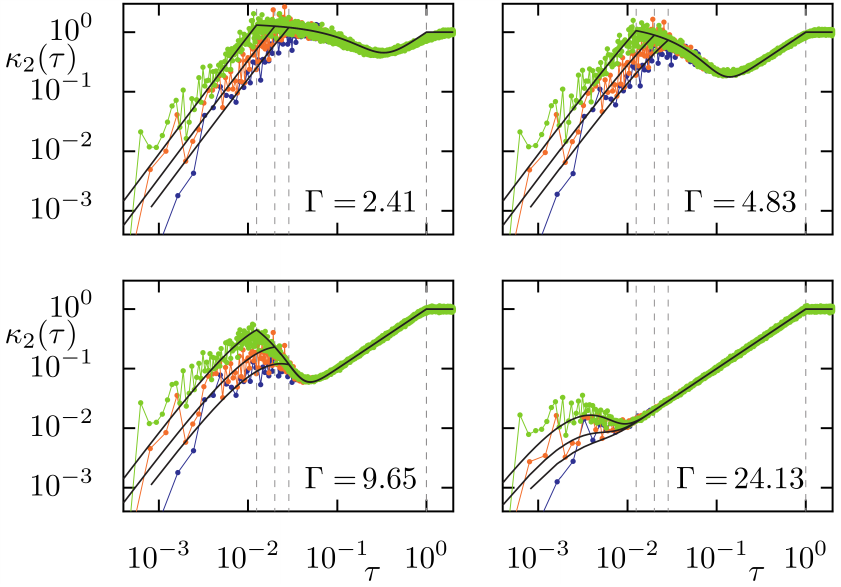}
    \caption{Rescaled second moment of the spectral form factor $\kappa_2(\tau)$ for the coupled kicked rotors for different $\Gamma$ (indicated in the individual panels) with $N=35$ (blue symbols, $10000$ realizations), $N=50$ (orange symbols, $10000$ realizations), and $N=80$ (green symbols, $1000$ realizations) in log-log scale with $N$ decreasing from top to bottom at small $\tau$.  The asymptotic result~\eqref{eq:moments_semiclassics_3} is depicted as black lines. Dashed gray lines correspond to the Heisenberg times $\tau_{\text{SH}}$ of the subsystems and of the bipartite system $\tau_{\text{H}}$.}
    \label{fig:sff_moment2_kicked_rotor_universality}
\end{figure}

\begin{figure}[]
    \centering
    \includegraphics[width=8.5cm]{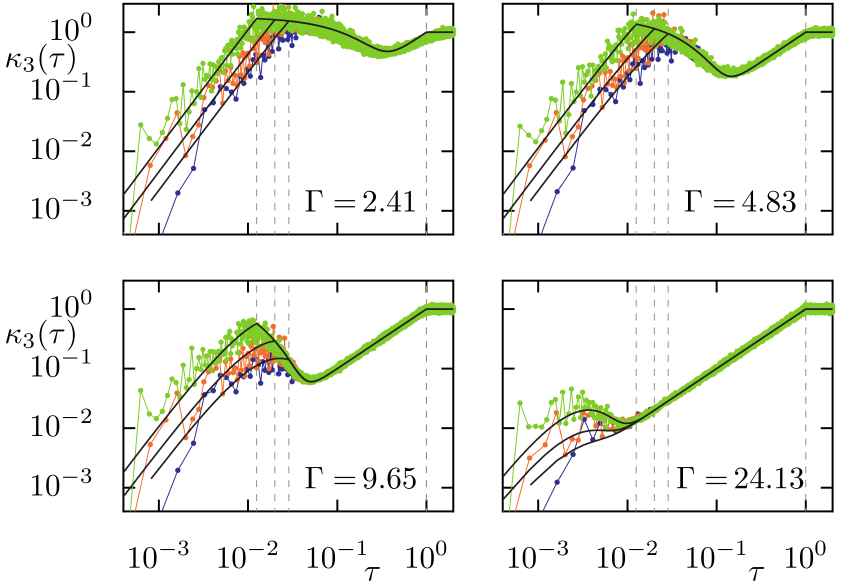}
    \caption{Same as Fig.~\ref{fig:sff_moment2_kicked_rotor_universality} but for the third moment of the spectral form factor. The asymptotic result (black lines) is given by Eq.~\eqref{eq:moments_semiclassics_4}.}
    \label{fig:sff_moment3_kicked_rotor_universality}
\end{figure}

To compare the numerical data with the asymptotic result~\eqref{eq:spectral_form_factor_semiclassics2} we introduce the following statistical model for the phases $\xi_{ij}$ entering the coupling $\U_{\text{c}}$.
We consider the phases to be of the form $\xi_{ij} = \cos\left(\eta_{ij}\right)$ with 
$\eta_{ij}$ i.i.d. random variables uniformly distributed on $[-\pi, \pi]$.
Hence the $\xi_{ij}$ are distributed with density
\begin{align}
f_\xi(x) = \frac{1}{\pi}\frac{1}{\sqrt{1 - x^2}}
\end{align}
for $x\in (-1, 1)$.
Consequently, the $\xi_{ij}$ have mean $0$ and variance $\sigma^2=1/2$.
Their characteristic function is given by the zeroth order Bessel function $J_0$.
Moreover, we identify the effective coupling strength as $\epsilon = \gamma N /(2\pi)$ leading to $\chi(\epsilon) = J_0(\gamma N /(2\pi))$ in Eq.~\eqref{eq:spectral_form_factor_semiclassics2}.
The resulting asymptotic result for the spectral form factor is depicted in 
Fig.~\ref{fig:sff_kicked_rotor_N50} by black lines.
It gives a good description of the average behavior of the spectral form factor with slightly worse accuracy compared to the RMTE.
The asymptotic result, however, does not take the small scale oscillations into account.
Moreover, the overall agreement of Eq.~\eqref{eq:spectral_form_factor_semiclassics2} with the numerically obtained spectral form factor implies universality of the latter. 
That is for $\tau>\tau_{\text{SH}}$ the spectral form factor of the coupled kicked rotors depends only on the scaling parameter $\Gamma$.
This is confirmed in Fig.~\ref{fig:sff_kicked_rotor_universality}, where the spectral form factor is shown for fixed values of $\Gamma$ but different $N$.
For higher moments of the spectral form factor we obtain qualitatively similar behavior, see Fig.~\ref{fig:sff_moment23_kicked_rotor_N50} for the second and third moment.
There we find good agreement between Eqs.~\eqref{eq:moments_semiclassics_3}~and~\eqref{eq:moments_semiclassics_4} and the numerically obtained moments for the second and third moment, respectively.
The implied universal dependence of the second and third moment on the scaling parameter $\Gamma$ is confirmed in Figs.~\ref{fig:sff_moment2_kicked_rotor_universality}~and~\ref{fig:sff_moment3_kicked_rotor_universality}.

Given the accuracy of the description by the asymptotic result Eq.~\eqref{eq:spectral_form_factor_semiclassics2} the Thouless time can be computed from  Eq.~\eqref{eq:Ehrenfest_time} for not to strong coupling $\gamma < 0.05$.
In contrast we might estimate the Ehrenfest time $t_{\text{E}}$ in the coupled kicked rotors
as follows: 
For weak coupling it is reasonable to assume, that the divergence of initially nearby trajectories predominantly takes place in the individual degrees of freedom.
Hence the coupled system has two positive Lyuapunov exponents 
$\lambda_i \approx \ln(k_i/2)$ \cite{Chi1979} each given by that of the subsystem $i \in \{A, B\}$, which in the present situation are approximately equal.
We confirm this argument by a numerical estimate of the largest Lyuapunov exponent of the coupled system (not shown).
This yields $t_{\text{E}} \approx \ln(N)/(2\ln(k_\ua k_\ub/4))$, which is much smaller than $N$.
Hence for weak enough coupling Ehrenfest and Thouless time do not agree in the coupled kicked rotors for the system sizes $N$ considered here and Ehrenfest time provides yet another relevant time scale.
In the semiclassical limit $N \to \infty$ at fixed coupling also $\Gamma \to \infty$ and the estimate Eq.~\eqref{eq:Ehrenfest_time} of the Thouless time is not valid anymore. 
Thus the above arguments still allow for Ehrenfest and Thouless time to approach each other in the semiclassical limit.

The presence of an underlying classical system provides also the possibility of applying the semiclassical description of the spectral form factor pioneered in Ref.~\cite{Gut1990,SieRic2001,Sie2002} and completed in Ref.~\cite{MueHeuBraHaaAlt2004} based on correlated periodic orbits.
Even though such a true semiclassical describtion is not attempted here, it would be interesting to compare it with the large $N$ asymptotics discussed above.
For instance semiclassical techniques might explain systematic fluctuations of the spectral form factor at small times and shed further light on the observed universal dependence on a single scaling parameter.

\section{Towards Many-Body Systems}
\label{sec:ext_rmte}

In the following we aim to extend our results to the many-body setting with an arbitrary number of subsystems. 
To this end we first introduce an extended version of the RMTE and subsequently generalize the computation of the spectral form factor. 
We will again derive its asymptotic form for large $N$ given as a convex combination of the uncoupled/non-interacting cases and the full random matrix result.
Moreover, we again obtain its universal dependence on a single scaling 
parameter $\Gamma$ which for the extended version additionally depends on the 
number of subsystems.

A natural extension of the bipartite RMTE~\eqref{eq:rmte} is given by 
 \begin{align}
    \U = \U_\uc(\epsilon) \left(\U_1 \otimes \U_2 \otimes \cdots \otimes \U_L 
    \right)
    \label{eq:ext_rmte}
 \end{align}
built from $L$ local subsystems, each of which is modeled by an independent unitary from the CUE$(N)$.
We keep the form of the coupling $\U_\uc(\epsilon)$ as a diagonal matrix with random phases, whose matrix elements in the computational basis read
\begin{align}
\bra{I}\U_{\uc}(\epsilon)\ket{J} =  
\delta_{IJ}\exp(\ui \epsilon \xi_{I}),
\label{eq:ext_coupling}
\end{align}
where we use multi-indices $I = (i_1,i_2,\ldots,i_L)$ for notational convenience.
The $\xi_I$ are again i.i.d. random variables with mean zero and finite variance $\sigma^2$.
We refer to $\U_\uc(\epsilon)$ and $\epsilon$ as interaction and interaction strength in the following.
In contrast to the random phase circuit of Ref.~\cite{ChaDeCha2018} we do not impose any spatial locality structure on the interaction and hence one might think of it as modeling an all-to-all or long-range interaction.
Nevertheless, the tensor-product structure of the subsystems induces a notion of locality in Hilbert space, which for the extended RMTE is the tensor product 
$\mathcal{H}=\left(\mathds{C}^N\right)^{\otimes L} = \mathds{C}^{N^L}$ and is of dimension $N^L$.

As is the case in the bipartite setting, we label the eigenphases defined by the eigenvalue equation
\begin{align}
    \U \ket{\Phi_I} = \ue^{\ui \varphi_I} \ket{\Phi_I}
\end{align}
by multi-indices as well for later convenience.
This is again motivated by the non-interacting case $\epsilon=0$, where eigenstates 
$\ket{\Phi_I}=\ket{\Theta_I}=\ket{\vartheta_{i_1}^{(1)}}\otimes \cdots \otimes \ket{\vartheta_{i_L}^{(L)}}$ are of product form and the eigenphases are the corresponding sums of individual eigenphases 
$\varphi_I(\epsilon=0) = \vartheta_{I} = \vartheta_{i_1}^{(1)} + \ldots + 
\vartheta_{i_L}^{(L)}$.

Taking the dimensionality of the underlying Hilbert space into account, the spectral form factor can now be written as 
\begin{align}
K(t) = \big\langle | \text{tr}\left(\U^t\right) |^2 \big\rangle  - N^{2L}\delta_{t0}
\label{eq:ext_spectral_form_factor_trace}
\end{align}
and can be computed in the semiclassical limit $N \to \infty$ by the same techniques as described in Sec.~\ref{sec:sff_semiclassics}.
More precisely the averages over the $L$ independent CUE$(N)$ and the phases factorizes.
Each individual average over the local Haar random unitaries $\U_i$ in leading 
order yields Eq.~\eqref{eq:haar_average_local}.
The spectral form factor then becomes 
\begin{align}
K(t) = N^{-Lt}\sum_{\mathbf{I}}\sum_{r_1,\ldots,r_L=0}^t \Big\langle \ue^{\ui \epsilon \theta\left(\mathbf{I}, (\eta_{r_1},\ldots, \eta_{r_L})\right)}\Big\rangle,
\label{eq:sff_ext_rmte1}
\end{align}
where we introduce the integer matrix $\mathbf{I} = \left(i_{ns}\right)_{ns} = \left(i_{s}^{(n)}\right)_{ns} \in \{1,\ldots,L\}\times \{1,\ldots,t\}$.
That is the first (upper) index $n$ labels the subsystem and the second (lower) index $s$ labels time.
That is $\mathbf{I}$ labels the canonical product basis in $\mathcal{H}^{\otimes t}$.
We denote the columns of this matrix by $I_s = \left(i_s^{(n)}\right)_n$.
Moreover, as in the bipartite case, we can restrict the sum in Eq.~\eqref{eq:sff_ext_rmte1} to those $\mathbf{I}$ for which all entries within a row are pairwise distinct
as these includes asymptotically all states in the product basis as $N \to \infty$.
For $L$-tuples of permutations $\boldsymbol{\mu}=\left(\mu_1,\ldots,\mu_L\right) \in S_t^L$ we write $\boldsymbol{\mu}\left(I\right) = \left(i_{\mu_n^{-1}(s)}^{(n)}\right)_{ns}$ for the matrix with permuted entries and $\boldsymbol{\mu}\left(I\right)_s$ for its columns.
That is $\mu$ permutes the entries in the $n$-th row by the permutation $\mu_n$.
With this notation the  generalization of Eq.~\eqref{eq:phases_for_averaging}
reads
\begin{align}
    \theta(\mathbf{I},\boldsymbol{\mu}) =  \sum_{s=1}^{t} \xi_{I_s} - \xi_{\boldsymbol{\mu}\left(I\right)_s}.
    \label{eq:theta_extended}
\end{align}
Repeating the argument which leads to Eq.~\eqref{eq:phases_averaged} we find
\begin{align}
\Big\langle \ue^{\ui \epsilon \mathbf{\theta}\left(\mathbf{I}, (\eta_{r_1},\ldots, \eta_{r_L})\right)}\Big\rangle = \delta_{r_1,\ldots,r_L} + (1-\delta_{r_1,\ldots,r_L})|\chi(\epsilon)|^{2t},
\label{eq:ext_phases_averaged}
\end{align}
where $\delta_{r_1,\ldots,r_L}=1$ when $r_1=r_2=\cdots=r_L$ and $0$ otherwise.
Inserting this into Eq.~\eqref{eq:sff_ext_rmte1} and noting that the sum over $\mathbf{I}$ runs asymptotically over $N^{Lt}$ states, which cancels the prefactor, we arrive at
\begin{align}
K(t) = |\chi(\epsilon)|^{2t}t^L + \left(1 - |\chi(\epsilon)|^{2t}\right) t.
\label{eq:sff_ext_rmte2}
\end{align}
Again this is exact as $N \to \infty$ and gives the leading contribution for $t< t_{\text{SH}}$ at large but finite $N$.
In the same fashion as for the bipartite case the above naturally extends to 
\begin{align}
K(t) = |\chi(\epsilon)|^{2t}K_N(t)^L + \left(1 - |\chi(\epsilon)|^{2t}\right) K_{N^L}(t),
\label{eq:sff_ext_rmte3}
\end{align}
where we again explicitly include the plateaus of $K_N(t)=N$ for $t>t_{\text{SH}}=N$ and $K_{N^L}=N^L$ for $t>t_{\text{H}}=N^L$, i.e., after the respective Heisenberg times.
The above reduces to Eq.~\eqref{eq:spectral_form_factor_semiclassics2} in the bipartite case $L=2$ and shows qualitatively very similar behavior.
In fact, in the non-interacting case, where $|\chi(\epsilon)|=1$, Eq.~\eqref{eq:sff_ext_rmte3} reduces to the factorized spectral form factor $K(t)=K_N(t)^L$ of $L$ independent CUE$(N)$ matrices.
Again, this is a consequence of Eq.~\eqref{eq:ext_spectral_form_factor_trace} and the multiplicativity of the trace, $\text{tr}\left(\bigotimes_iU_i\right) = \prod_i \text{tr}\left(U_i\right)$, see also Ref.~\cite{AgrPan2021}.
For not too large interaction, the initial growth $K(t)\sim t^L$ is still governed by this factorized behavior.
This coincides with what is observed in typical many-body systems including those with short-range interactions.
In contrast, at late times all the terms $|\chi(\epsilon)|^{2t}$ have decayed and the spectral form factor is given by the full random matrix result $K(t) \sim K_{N^L}(t)$.
To summarize, Eq.~\eqref{eq:sff_ext_rmte3} signals a competition between the locality in the system's Hilbert space induced by the tensor product structure and the global interaction.
This is encoded in the spectral form factor being again a time dependent convex combination of the two limiting cases of the non-interacting and the strongly interacting system.

Before discussing and comparing the extrapolated result with numerical data let us again introduce the rescaled spectral form factor by measuring both spectral form factor and time in units of Heisenberg time $t_{\text{H}} = N^L$.
This reads
\begin{align}
\kappa(\tau) = K(t)/N^L,
\label{eq:rescaled_ext_sff}
\end{align}
where $\tau = t/N^L$.
Consequently, the Heisenberg time of the full system again reads $\tau_{\text{H}}=1$ whereas the Heisenberg times of the subsystems become $\tau_{\text{SH}} = N^{L-1}$.
After rescaling and for times $\tau > \tau_{\text{SH}}$ the remaining $N$ dependence is again implicitly contained in the factors $|\chi(\epsilon)|^{2t}$ for which we might repeat the application of the central limit theorem.
In particular Eq.~\eqref{eq:central_limit_theorem2} is valid also in the extended setting but the universal scaling parameter now becomes
\begin{align}
    \Gamma = \sigma \epsilon N^{L/2},
\end{align}
which reduces to Eq.~\eqref{eq:scaling_parameter} in the bipartite case $L=2$. \\

\begin{figure}[]
    \centering
    \includegraphics[width=8.5cm]{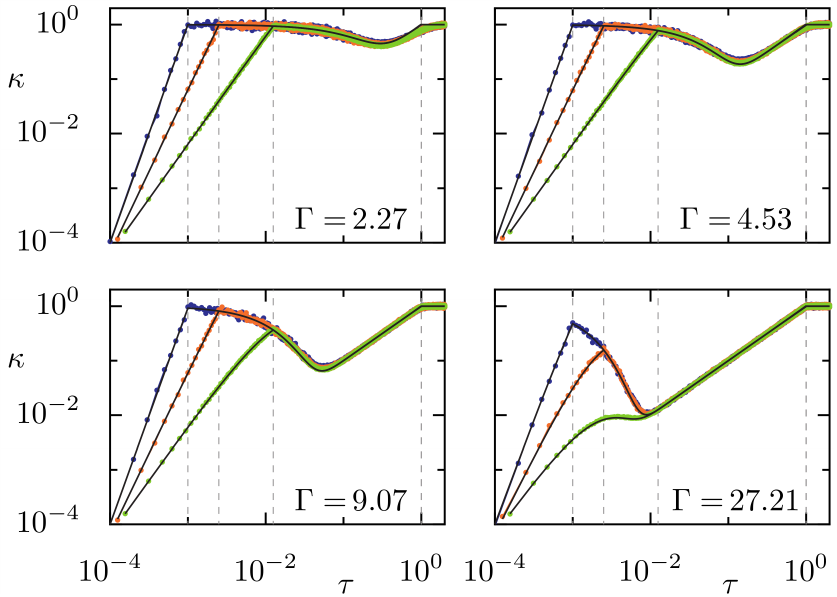}
    \caption{Rescaled spectral form factor $\kappa(\tau)$ for the extended RMTE with uniformly distributed phases for different $\Gamma$ (indicated in the individual panels) with $(N,L)=(10,4)$ (blue symbols, $4000$ realizations), $(N,L)=(20,3)$ (orange symbols, $4000$ realizations), and $(N,L)=(80,2)$ (green symbols, $6000$ realizations) in log-log scale with $N$ increasing from top to bottom at small $\tau$.  The asymptotic result~\eqref{eq:sff_ext_rmte3} is depicted as black lines. Dashed gray lines correspond to the Heisenberg times $\tau_{\text{SH}}$ of the subsystems and of the full system $\tau_{\text{H}}$.}
    \label{fig:sff_ext_rmte}
\end{figure}

As our arguments rely on $N$ being large, numerical investigations are limited to only a few subsystems.
That is, the validity of the extrapolated result can only be confirmed for few-body systems.
However, as none of our arguments relies on $L$ being small, we expect our results to apply also in the many-body setting of large $L$ as long as $N$ is sufficiently large.
In Fig.~\ref{fig:sff_ext_rmte} we depict the rescaled spectral form factor for few-particle systems with $L=3,4$ as well as the corresponding data for a bipartite system for comparison. 
We find qualitative similar behavior for all $L$ except for the initial growth of the spectral form factor as $\kappa(\tau)\sim \tau^L$. 
Otherwise the discussion from the bipartite case also applies for the extended random matrix transition ensemble.
In particular we observe the same universal dependence of the spectral form factor on the scaling parameter $\Gamma$.

Given the universal dependence on $\Gamma$ the Thouless time $t_{\text{Th}}$ for intermediate interaction strength, i.e., such that $t_{\text{SH}} < t_{\text{Th}} < t_{\text{H}}$ is still described by Eq.~\eqref{eq:thouless_time_analytic}.
In contrast, adapting the arguments leading to Eq.~\eqref{eq:Ehrenfest_time} to the extended RMTE yields
\begin{align}
    t_{\text{Th}} = \frac{L\ln(N)}{2|\ln(|\chi(\epsilon)|)|}.
\end{align}
Hence we find the Thouless time to scale linear with system size, i.e., the number of subsystems.
This is notably different from the scaling observed in chaotic systems with local interactions.
For instance logarithmic scaling is found both in the large $N$ case \cite{ChaDeCha2018} and in qubit systems \cite{KosLjuPro2018,ChaDeCha2018}, whereas the presence of $\text{U}(1)$ symmetry yields quadratic scaling \cite{RoyPro2020,RoyMisPro2022}.
Moreover, also exponential \cite{SchTorSan2019} as well as subdiffusive scaling \cite{ColLuiKhaDeT2022} has been ovserved. In contrast dual unitary quantum circuits even exhibit zero Thouless time in the thermodynamic limit \cite{BerKosPro2018, BerKosPro2021}.

Using similar arguments as above and in Sec.~\ref{sec:sff_moments} also the moments of the spectral form factor in the extended RMTE could in principle be obtained.
We do not attempt to give an exact description even for short times, but nevertheless provide a sketch of the derivation which yields some qualitative aspects of the moments, which we complement by numerical results.
When deriving the moments of the SFF in the extended RMTE the average over the independent CUE$(N)$ matrices proceeds in the same way as in Sec.~\ref{sec:sff_moments}.
In contrast, the average over the phases $\xi_I$ is more involved.
First, we note, that the moments of the spectral form factor asympotically read
\begin{align}
    K_m(t)=N^{-Lmt}\sum_{\mathbf{I}}\sum_{\boldsymbol{\mu} \in G_m^L}\Big\langle \ue^{\ui \epsilon \theta\left(\mathbf{I}, \boldsymbol{\mu} \right)} \Big\rangle 
    \label{eq:moments_extended_rmte1}
\end{align}
as $N \to \infty$.
Here, $\mathbf{I}$ now labels the product basis in $\mathcal{H}^{\otimes mt}$ and the sum in Eq.~\eqref{eq:theta_extended} runs up to $mt$, with the action of a permutation $\boldsymbol{\mu}=\left(\mu_1,\ldots,\mu_L\right) \in S_{mt}^L$ defined as in the case of the spectral form factor, i.e., $m=1$.
Similarly, for asymptotically all states $\mathbf{I}$ the average over the phases does not depend on the state and the corresponding sum gives a factor $N^{Lmt}$ thereby canceling the prefactor in Eq.~\eqref{eq:moments_extended_rmte1}.
We are thus left with computing the sum over $\boldsymbol{\mu} $.

To this end for $\boldsymbol{\mu} \in G_m^L$ we denote by $c(\boldsymbol{\mu})$ the number of points $x$ on which all the $\mu_i$ agree, i.e, $\mu_1(x) =  \ldots = \mu_L(x)$.
The average over the phases than gives
\begin{align}
\Big\langle \ue^{\ui \epsilon \theta (\mathbf{I}, \boldsymbol{\mu})} \Big\rangle = |\chi(\epsilon)|^{2(tm - c(\boldsymbol{\mu}))}
\label{eq:phase_average_ext}
\end{align}
independent from asymptotically all states $\mathbf{I}$.
Equivalently to the above definition, $c(\boldsymbol{\mu})$ is the number of common fixed points of the $L-1$ permutations $\mu^{-1}_1\mu_2,\ldots,\mu^{-1}_1\mu_2$.
A similar argument as in the bipartite case implies that $c(\boldsymbol{\mu})=kt$ with $k \in \{0,1,\ldots,m\}$.
We denote by $A_k(t)$ the number of $(L-1)$-tuples of permutations, i.e., elements of $G_m^{L-1}$, with exactly $kt$ common fixed points. 
By a change of variables the sum over $\mu_1$ trivializes and gives a total factor of $|G_m|=m!t^m$.
The spectral form factor therefor can be rewritten as 
\begin{align}
K_m(t)=m!t^m \sum_{k=0}^m A_k(t)|\chi(\epsilon)|^{2t(m - k)}
\label{eq:moments_extended_rmte2}
\end{align}
just as in the bipartite case.
The combinatorical factors $A_k(t)$ are polynomials of degree at most $m(L-1)$.
Their computation is considerably more involved than in the bipartite case and is not attempted here.

However, some qualitative features of the moments of the spectral form factor can still be read off from Eq.~\eqref{eq:moments_extended_rmte2}.
First, one has $\sum_k A_k(t) = |G_m|^{L-1} = \left(m!t^m\right)^{L-1}$ which in the non-interacting case, $|\chi(\epsilon)|=1$, implies $K_m(t)=\left(m!t^m\right)^{L}$, i.e., the expected factorization of the moments $K_m(t)=\left[K_{N,m}(t)\right]^L$.
For small interaction we therefore expect an initial growth $K_m(t) \propto t^{mL}$.
Second, in the interacting case all terms with $k\neq m$ decay exponentially and for $t \gtrsim t_{\text{Th}}$ only the term for $m=k$ survives.
As the latter is given by $A_m(t)=1$ the spectral form factor reduces to the random matrix result $K_m(t)=K_{N^L, m}(t) = m!t^m$ and hence yields an exponential distribution.
Additionally, when viewing the combinatorical factors $A_k(t)$ as a polynomial in $t$ the coefficients will depend on $L$, which will lead to the moments of the spectral form factor to depend both on $\Gamma$ and $L$ also for times $t>t_{\text{SH}}$.

\begin{figure}[]
    \centering
    \includegraphics[width=8.5cm]{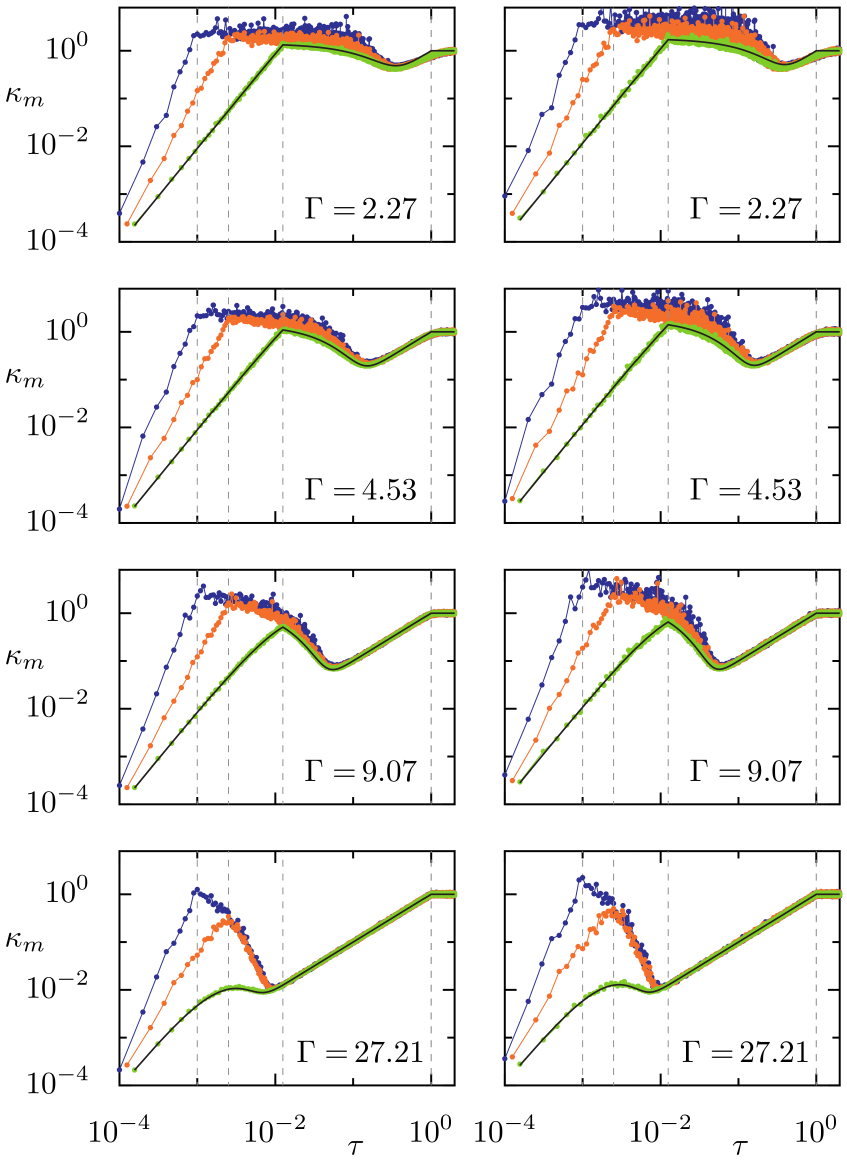}
    \caption{Same as Fig.~\ref{fig:sff_ext_rmte} but for the 
        rescaled second ($m=2$, left) and third ($m=3$, right) moment of the spectral form factor $\kappa_m(\tau)$. The asymptotic results~\eqref{eq:moments_semiclassics_3} and ~\eqref{eq:moments_semiclassics_4} are depicted as black lines for the bipartite system. Dashed gray lines correspond to the Heisenberg times $\tau_{\text{SH}}$ of the subsystems and of the full system $\tau_{\text{H}}$.}
    \label{fig:sff_moments23_ext_rmte}
\end{figure}

We confirm the above considerations numerically in Fig.~\ref{fig:sff_moments23_ext_rmte} for the second and third rescaled moment
\begin{align}
\kappa_m(\tau) = \frac{1}{N^L}\left(\frac{K_m(t)}{m!} \right)^{1/m}.
\label{eq:moments_ext_sff_rescaled}
\end{align}
For comparison we also show the bipartite case there.
In general the rescaled moments follow the same phenomenology as the spectral form factor.
The most notable difference is the failure of the universal dependence on $\Gamma$ between systems with different numbers of subsystems $L$.
In fact for rescaled times $\tau_{\text{SH}} < \tau \lesssim \tau_{\text{Th}}$ the moments do depend also on $L$ with larger $L$ leading to larger $\kappa_m(\tau)$.
Only for times approximately larger than the Thouless time, $\tau \gtrsim \tau_{\text{Th}}$ the moments follow the exponential distribution of the CUE$(N^L)$ moments expressed as $\kappa_m(\tau) = \min\{\tau, 1\}$ independently from $L$.

\section{Regularized Perturbation Theory \label{sec:perturbation_theory}}

In this section we consider the spectral form factor of the extended RMTE at weak interaction strength and small $\Gamma$ by using the 
conventional Rayleigh-Schr\"odinger perturbation theory.
In fact, it turns out that instead of $\Gamma$ the universal behavior of the spectral form factor is governed by a so-called transition parameter $\Lambda$ \cite{SriTomLakKetBae2016}, which naturally arises in the perturbative approach.
However, in the perturbative regime, both parameters agree in leading order.
The perturbative approach, despite being restricted to weak interactions covers 
arbitrary long times and has been applied successfully for e.g. the entanglement 
dynamics after a quench \cite{PulLakSriBaeTom2020} as well as static properties
\cite{SriTomLakKetBae2016,LakSriKetBaeTom2016,TomLakSriBae2018,HerKieFriBae2020}.
The approach is based 
on the pertubative expansion of the eigenphases $\varphi_I$.
In fact knowledge of the eigenphases is sufficient for computing the spectral form factor as
 Eq.~\eqref{eq:ext_spectral_form_factor_trace} for $t > 0$ can be written as 
\begin{align}
K(t) = \Big\langle\sum_{I, K} \ue^{\ui (\varphi_I-\varphi_K) t}\Big\rangle
\label{eq:sff_from_eigenphases}
\end{align}
by computing the trace in the eigenbasis of the interacting system.
For later analytical treatment it is convenient to rewrite this in terms of the rescaled spectral form factor, Eq.~\eqref{eq:rescaled_ext_sff}, via
\begin{align}
\kappa(t) = 1 + \frac{1}{N^L} \Big\langle \sum_{I \neq K} \cos( \Delta \varphi_{I, K} t) 
\Big\rangle. \label{eq:sff_from_eigenphase_diffs}
\end{align}
Here we denote the differences of eigenphases by $\Delta \varphi_{I, K} = \varphi_I - 
\varphi_K$ .

To obtain the perturbative expansion of the eigenphases and their differences we write $\U_\uc(\epsilon) = \exp(\ui \epsilon V)$ with $V$ the Hermitian diagonal matrix with entries $\bra{I}V\ket{J} = \delta_{IJ}\xi_I$ in the canonical product basis.
In second order in $\epsilon$ we obtain \cite{TomLakSriBae2018}
\begin{align}
\varphi_{I} &= \vartheta_{I} 
+ \epsilon \bra{\Theta_I} V \ket{\Theta_I} + \epsilon^2 \sum_{I \neq J} \frac{|\bra{\Theta_I}V\ket{\Theta_J}^2|}{\vartheta_{I} - \vartheta_{J}}
\label{eq:eigenphases_pert}
\end{align}
where, as introduced above, $\ket{\Theta_I}$ and  $\vartheta_{I}$
are the eigenvectors and eigenphases of the non-interacting system.
Therefore, the phase differences 
$\Delta \varphi_{I, K}$  read
\begin{align} \label{eq:delta_phi_rough_expr}
\Delta \varphi_{I, K} & = 
\Delta \vartheta_{I, K} 
+ \epsilon T^{(1)}_{I,K}
+ 2 \epsilon^2 \frac{|\bra{\Theta_I}{V}\ket{\Theta_K}|^2}
{\Delta \vartheta_{I, K} } 
+ \epsilon^2  T^{(2)}_{I,K}.
\end{align}
Here  $\Delta \theta_{I, K} = 
\vartheta_{I} - \vartheta_{K}$ denotes the unperturbed level difference.
Moreover, the two terms 
\begin{align}
T^{(1)}_{I,K} &= \matrixel{\Theta_I}{V}{\Theta_I} - \matrixel{\Theta_K}{V}{\Theta_K}, \\
T^{(2)}_{I,K} &= \sum_{I \neq J\neq K}\left(
\frac{|\bra{\Theta_I}{V}\ket{\Theta_J}|^2}{\Delta\vartheta_{I,J}} 
- \frac{|\bra{\Theta_K}{V}\ket{\Theta_J}|^2}{\Delta\vartheta_{K,J}} \right)
\end{align}
do not contribute to the phase difference after performing the average over the extended RMTE.
More precisely, this is justified as the first order term $T^{(1)}_{I,K}$ introduces 
just an overall phase shift \cite{PulLakSriBaeTom2020} which does not contribute to the spectral form factor.
The second order term $T^{(2)}_{I,K}$
contains only contributions which do not connect both level simultaneously, such that each 
yields a random and small background which can also be neglected upon averaging \cite{PulLakSriBaeTom2020}. 

For further analytical treatment it is convenient to rescale both the phase differences and the matrix elements of the interaction $V$.
In particular we write $\Delta \theta_{I, K} = D s_{I,K}$, where $D=2\pi/N^L$ is the mean level spacing, such that $s_{I,K}$ has unit mean spacing.
Moreover, we rewrite the matrix elements 
$|\matrixel{\Theta_I}{V}{\Theta_K}|^2 = 
\nu^2 w_{I, K}$ where $\nu^2$ is the mean of the square modulus of the 
off diagonal of the interaction in the eigenbasis $\ket{\Theta_I}$ of the non-interacting system.
Inserting these substitutions into the perturbative expansion~\eqref{eq:delta_phi_rough_expr} yields
\begin{align}
\Delta \varphi_{I,K} = Ds_{I,K}\left(1 + \frac{2\Lambda w_{I,K}}{s_{I,K}^2}\right).
\label{eq:phase_difference_substitution}
\end{align}
Here,
\begin{align}
    \Lambda = \frac{\epsilon^2\nu^2}{D^2}
\end{align}
denotes so-called transition parameter, which originally was introduced in the bipartite RMTE \cite{SriTomLakKetBae2016} and universally governs the transition of eigenstate properties in the bipartite case \cite{SriTomLakKetBae2016,LakSriKetBaeTom2016,TomLakSriBae2018,HerKieFriBae2020}.
For the extended RMTE the transition parameter reads 
\begin{align}
\Lambda = \frac{N^L}{4\pi^2}\left(1 - |\chi(\epsilon)|^2\right),
\label{eq:transition_parameter}
\end{align}
for large $N$, see App.~\ref{App:PerturbationTheoryTransitionParameter} for details.
Here, $\chi(\epsilon)$ again denotes the characteristic function of the distribution of the phases $\xi_I$.
Expanding the latter up to second order in $\epsilon$ and minding that we assume the distribution of the
phases $\xi_I$ to have zero mean we obtain
\begin{align}
\Lambda = \frac{\sigma^2\epsilon^2N^L}{4\pi^2} = \frac{\Gamma^2}{4\pi^2}.
\label{eq:lambda_vs_gamma}
\end{align}
This clarifies the connection between the transition parameter $\Lambda$, which 
naturally arises in the above argumentation, and the scaling parameter $\Gamma$ 
for small interaction strength $\epsilon$.
In the following we base our discussion on $\Lambda$ nevertheless but given the 
above relation between the two parameters properties depending on $\Lambda$ 
translate to a dependence on $\Gamma$ and vice versa.
For uniformly distributed phases used for numerical computations one has $\Lambda = N^L\epsilon^2/12$.

Coming back to the evaluation of Eq.~\eqref{eq:sff_from_eigenphase_diffs} we proceed by replacing the sum over
the eigenphases $\sum_{I \neq K}\left(\cdot\right)$ by an integral 
$\int\! \ud s \int\! \ud w \left(\cdot\right)R(s, w)$ 
 over random variables $s \in (-\infty, \infty)$ and $w \in [0, \infty)$ with distribution
\begin{align}
R(s, w) = \sum_{I,K} \delta(s - s_{I, K})\delta(w - w_{I,K}).
\end{align}
The ensemble average over the RMTE now reduces to an average over the above probability distribution.
Assuming independence of eigenvectors and eigenphases for the non-interacting extended RMTE, i.e., for the $L$-fold tensor product of independent CUE$(N)$ the distribution factorizes as \cite{SriTomLakKetBae2016}
\begin{align}
\langle R(s, w) \rangle = R(s) \ue^{-w}
\end{align}
Here, the matrix element distribution follows uncorrelated Poissonian statistics 
and 
the two-point level distribution $R(s)$ for the uncoupled case is given by $R(s) = 
1$ in the limit $N \to \infty$ \cite{Meh1991}.
Note that $R(s)$ is conceptually similar to $r(\omega)$ defined by Eq.~\eqref{eq:spectral_twopoint_function} as $N \to \infty$ after proper rescaling.
However, in this limit only the non-connected part is non-zero.
Furthermore we emphasize that $R(s)=1$ at finite $N$ fails to capture correlations in the spectrum of the non-interacting 
RMTE on energy scales larger than the mean level spacing of the subsystems.
Consequently, the perturbative approach fails to capture the initial dynamics of the spectral form factor
$K(t) \sim t^L$ for times $t < t_{\text{Th}}$.
This regime, however, is well described by Eq.~\eqref{eq:spectral_form_factor_semiclassics2} whereas the perturbative approach describes times $t>t_{\text{Th}}$, which are not accurately captured by Eq.~\eqref{eq:spectral_form_factor_semiclassics2}.
We henceforth focus only on the latter time regime in the remainder of the section.

To evaluate the integral over the random variables $s$ and $w$ introduced above
we note, that the integral diverges due to the singular behavior of Eq.~\eqref{eq:phase_difference_substitution} as $s \to 0$.
Physically this is caused by the breakdown of non-degenerate perturbation theory leading to Eq.~\eqref{eq:delta_phi_rough_expr} in case of degenerate levels.
For almost degeneracies in the spectrum of the non-interacting RMTE degenerate perturbation theory within an effective two-level systems suggests the regularization 
\cite{CerTom2003,TomLakSriBae2018,PulLakSriBaeTom2020}
\begin{align}
\Delta \varphi_{I, K} = D\sqrt{s^2 + 4 \Lambda w}.
\end{align}
This renders the integrals finite as it removes the singularity at $s=0$, while coinciding with Eq.~\eqref{eq:phase_difference_substitution} at large spacing $s$.

\begin{figure}[]
    \centering
    \includegraphics[width=8.5cm]{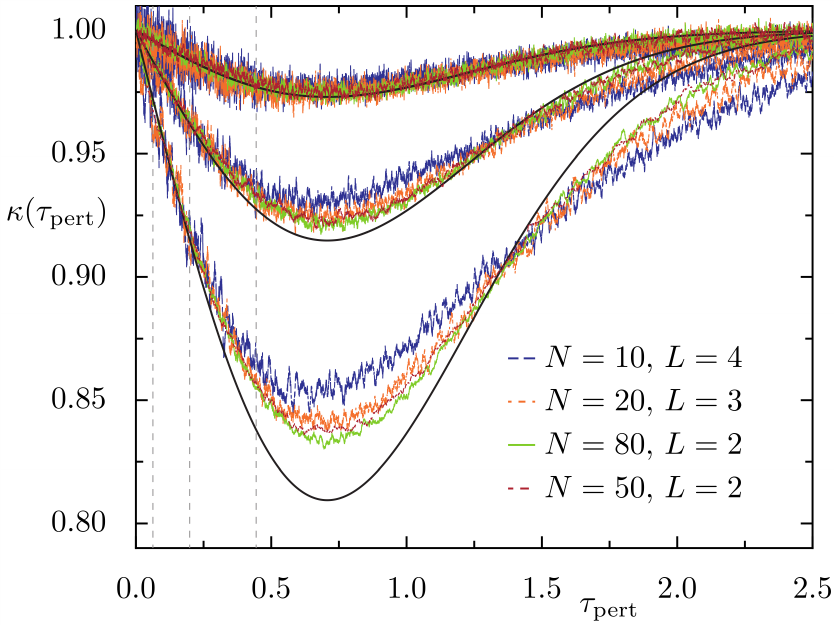}
    \caption{Rescaled spectral form factor $\kappa(\tilde{\tau})$ for the RMTE for
        various combinations of $N$, $L$ (see legend) and $\epsilon$
        corresponding to 
        three different values of $\Lambda=0.001,\,0.005,\,0.010$ (top to bottom). Thin colored 
        lines correspond to numerical data obtained from $2000$ realizations of the RMTE 
        with uniformly distributed phases $\xi_{I}$. We show those data for times $t>t_{\text{H}}$ and perform an additional moving time 
        average. The perturbative result~\eqref{eq:sff_rmte_perturbation_theory} is 
        depicted as thick black lines. Vertical dashed gray lines correspond to the Heisenberg times for the different system sizes. \label{fig:sff_rmte_perturbation_theory}}
\end{figure}

Putting all the above together we end up writing the second term in  Eq.~\eqref{eq:sff_from_eigenphase_diffs} as
\begin{align}
\mathcal{I} & = \frac{1}{N^L}\Big\langle \sum_{I \neq K}\cos\left(\Delta \varphi_{I, K}\right) \Big\rangle  \\
& =
\int\limits_{-\infty}^{\infty}\ud s \int\limits_{0}^{\infty} \ud w \cos\left(Dt\sqrt{s^2 + 4 \Lambda w}\right)\ue^{-w}. \label{eq:perturb_int}
\end{align} 
We evaluate this integrals in App.~\ref{App:PerturbationTheoryIntegral}.
The details of the computation suggest the introduction of a rescaled time variable $\tau_{\text{pert}}$
given by \cite{PulLakSriBaeTom2020}
\begin{align}
    \tau_{\text{pert}} = \sqrt{\Lambda} D t,
    \label{eq:tau_pert}
\end{align}
which differs from the variable $\tau$.
In terms of this perturbatively rescaled time variable the final result reads
\begin{align} \label{eq:sff_perturbativ_integral}
\mathcal{I} = -2 \pi \sqrt{\Lambda} \tau_\text{pert} \ue^{-\tau_\text{pert}^2}.
\end{align}
Hence we obtain the spectral form factor as 
\begin{align} 
\kappa(\tau_\text{pert}) = 1 - 2 \pi \sqrt{\Lambda} \tau_\text{pert} 
\ue^{-\tau_\text{pert}^2}.
\label{eq:sff_rmte_perturbation_theory}
\end{align}
This universally depends on the transition parameter $\Lambda$ only and captures the spectral form factor for physical times $t>t_{\text{SH}}$, corresponding to 
$\tau_{\text{pert}}>\sigma \epsilon N^{-L/2}$.
It gives an accurate description of numerical data for weak interactions and hence small transition parameters as it is confirmed by Fig.~\ref{fig:sff_rmte_perturbation_theory} for different combination of 
$N$, $L$ and $\epsilon$ all leading to the same transition parameter.
We find good agreement for essentially arbitrary long times, i.e., multiples of Heisenberg time, at very small transition parameter.
In contrast, the data shown for the largest transition parameter indicate the 
breakdown of the validity of the perturbative approach.
Despite Eq.~\eqref{eq:sff_rmte_perturbation_theory} becoming less accurate for 
larger $\Lambda$ the numerical data nevertheless confirm the universal 
dependence of the spectral form factor on $\Lambda$.
This is even the case for surprisingly small $N=10$ with deviations being of the expected order $1/N$.

Given the connection between the transition parameter $\Lambda$ and the scaling 
parameter $\Gamma$ highlighted in Eq~\eqref{eq:lambda_vs_gamma} this further 
confirms the universal dependence of the spectral form factor also on $\Gamma$ 
obtained by complementary methods.
Moreover, we might write the perturbative result also in terms of $\Gamma$ and $\tau$ which yields
\begin{align}
\kappa(\tau) = 1 - \Gamma^2 \tau \ue^{-(\Gamma \tau)^2}. \label{eq:sff_rmte_perturbation_theory_Gamma}
\end{align}
for $\tau>N^{-L + 1}$ and allows for a comparison with Eq.~\eqref{eq:sff_ext_rmte3}.
Noting that the latter, extrapolated result fails to capture the spectral form factor at long times for small $\Gamma$ we can only hope for both approaches to coincide for times $\tau = \tau_{\text{SH}} + \delta \tau$ for small $\delta \tau$.
This covers the times in which the spectral form factor starts to drop again for times slightly larger than the subsystems' Heisenberg times.
Indeed, the linearization of both approaches agree in leading order in $\Gamma$, i.e.,
\begin{align}
\kappa(\tau) \approx 1 -\Gamma^2 \delta \tau
\end{align}
indicating that both approaches are compatible in the regime where both apply.

Ultimately, let us briefly comment on the extension of the perturbative approach to higher moments of the spectral form factor in the bipartite or extended RMTE.
Doing so for the $m$-th moment requires to evaluate an expression similar to Eq.~\eqref{eq:sff_from_eigenphases} but with a $2m$ fold sum and with sums and differences of $2m$ eigenphases in the exponent. 
While the perturbative expansion~\eqref{eq:eigenphases_pert} as well as the subsequent substitutions still apply, replacing the sums with integrals over appropriate random variables becomes more involved.
For instance the relevant probability density for the second moment reads
\begin{align}
\langle R(s_1, w_1)R(s_2, w_2) \rangle = & R_3(s_1, s_2)\ue^{-w_1 -w_2} +  \nonumber \\
& \delta(s_1 - s_2)\delta(w_1 - w_2)R_2(s_1)\ue^{-w_1}.
\end{align}
where $R_k(s) = 1$ is the  $k$-point function of the level density. 
We do not attempt to solve the resulting integrals here.

\section{Conclusions}
\label{sec:conclusions}

We study spectral correlations in bipartite and few-body chaotic quantum systems in terms of the spectral form factor and its dependence on the interaction between the subsystems.
We find universal dependence of the spectral form factor and, in the bipartite case, its higher 
moments on a universal scaling parameter $\Gamma$.
This is derived for the extended random matrix transition ensemble which is tailored to 
describe the spectral properties of interacting chaotic few-body systems, for which the scaling parameter combines all the details of the system, i.e., system size, number of subsystems as well as  strength and statistics of the interaction.
The RMTE allows for computing the spectral form factor exactly in the semiclassical limit for times smaller than the subsystems' Heisenberg times.
These results permit an extrapolation to larger times as a time dependent convex combination of the non-interacting and the strongly interacting limit.
For sufficently large $N$ the extrapolated result is in  good agreement with numerically obtained data and gives insight in the origin of the observed universality.
Interestingly, $N=10$ seems to be large enough for our description to apply in the few-body setting.
A more rigorous derivation of the spectral form factor for times larger than 
the subsystems' Heisenberg times is out of scope of the large $N$ asymptotics 
exploited here, but might be accessible by supersymmetric field theoretic methods \cite{Zir1996,AltGnuHaaMic2015}.
Obtaining exact finite $N$ results would be of great interest, in particular in the many-body setting, for which our extrapolated results in the extended RMTE provide a first step.

In principle, the methods presented here should be applicable also to bipartite or multipartite systems with time reversal symmetry, for which, e.g., the subsystems are modeled by the circular orthogonal ensemble.
Also different statistical models for the coupling might be treatable.
Ultimately, the large $N$ analysis might apply also to different observables, i.e., eigenstate entanglement or entangling power.

We complement the large $N$ analysis with a perturbative treatment of the coupling in the extended RMTE 
using regularized Rayleigh-Sch\"odinger perturbation theory. 
We find universal dependence of the spectral form factor on a transition 
parameter $\Lambda$. It coincides with $\Gamma$ in the regime where both, 
the large $N$ asymptotics and the perturbative approach apply.
For small transition parameters, e.g.~for small couplings, the 
perturbative result describes numerical data well for all times larger 
than the Heisenberg time of the subsystems.
In particular the perturbative approach captures deviations from the full random matrix spectral form factor for times larger than the coupled system's Heisenberg times in the regime of the plateau.
An extension of the perturbative approach to higher moments might shed more light on the spectral form factor at small coupling, but is not attempted here.

Finally, we confirm that the large $N$ results accurately describe spectral correlations also in more realistic physical systems.
Within the paradigmatic coupled kicked rotors we study the spectral form factor and its moments and find them well described by the results obtained in the RMTE.
Moreover, the presence of an underlying classical system allows for comparing Thouless time, i.e., the time after which the spectral form factor follows the linear ramp obtained for $\text{CUE}(N^2)$, and Ehrenfest time, i.e., the time it takes for an initially localized wave packet to spread over the system.
At least for small coupling and finite $N$ we clearly demonstrate that both times are different.

The underlying classical system of the coupled kicked rotors further yields the 
possibility to explore spectral correlations by genuine semiclassical methods, 
e.g., in terms of correlated periodic orbits.
This might give further insight into the observed phenomena, such as the universal dependence on $\Gamma$, but is far beyond the scope of the present paper.

\section*{Acknowledgements}

We thank A. B\"{a}cker, P. Kos, F.~G.~Montoya, and T. Prosen for insightful discussions.
The work has been supported by Deutsche Forschungsgemeinschaft (DFG), Project No. 453812159 (FF) and Project No. 497038782 (MK).

\appendix

\section{Leading Contributions to Moments of the Spectral Form Factor}

In this appendix we characterize the permutations $\mu \in S_{mt}$ which determine the $m$-th moment of the spectral form factor as $N\to \infty$.
Moreover, we use this characterization to derive Eq.~\eqref{eq:A_kt}.

\subsection{Contributing Permutations}
\label{App:Permutations}

We first aim to find those permutations $\mu$ which when evaluating the Haar average over the subsystems $\U_\ua$ and $\U_\ub$ in terms of Weingarten functions give rise to a non-vanishing contribution.
That is, we identify those $\mu \in S_{mt}$ which are invariant under conjugation by $\eta_1^{\otimes m}$ as argued in Sec.~\ref{sec:sff_moments}, i.e., the solutions to 
\begin{align}
    \mu \eta_1^{\otimes m}  = \eta_1^{\otimes m} \mu.
    \label{appeq:leading_contribution_permutations}
\end{align}

To this end we introduce the following characterization of $\eta_1^{\otimes m} \in S_{mt}$. Each
$x \in \{0,1, \ldots, mt - 1\}$ has a unique representation $x=s + nt$ with $s \in \{0,1,\ldots,t-1\}$ and $n \in \{0,1,\ldots,m-1\}$.
This yields $\eta_1^{\otimes m}(x) = \eta_1(s) + nt = \left[(s+1)\!\!\mod t\right] + nt$.
Using this representation we show how the group $G_m=S_m \ltimes \langle \eta_1 \rangle^m$ is embedded in $S_{mt}$ and that each elements $\mu \in G_m$ solves Eq.~\eqref{appeq:leading_contribution_permutations}.
Subsequently we argue, why these are indeed all solutions.

An element $\mu \in G_m$ is of the form $\mu = \left(\rho, \eta_{r_0}, \eta_{r_2}, \ldots, \eta_{r_{m-1}}\right)$ with $\rho \in S_m$ and $r_i \in \{0, 1, \ldots, t-1\}$.
We define the action of $\mu$ on $x=s + nt$ by
\begin{align}
    \mu(x) = \eta_{r_n}(s) + \rho(n)t,
\end{align}
which allows to interpret $\mu$ as an element of $S_{mt}$.
More precisely, the above construction gives rise to an injective group homomorphism $G_m \to S_{mt}$ and hence $G_m$ can be identified with a subgroup of $S_{mt}$.
Moreover $\mu \in G_m$ solves Eq.~\eqref{appeq:leading_contribution_permutations} as
\begin{align}
\mu\eta_1^{\otimes m}(x) & = \mu\left(\eta_{1}(s) + nt\right) \\
& = \eta_{r_n}\eta_{1}(s) + \rho(n)t \\
& = \eta_{1}\eta_{r_n}(s) + \rho(n)t \\
& = \eta_1^{\otimes m}\left(\eta_{r_n}(s) + \rho(n)t\right) \\
& = \eta_1^{\otimes m}\mu(x).
\end{align}

Now let us assume $\mu \in S_{mt}$ is a solution to  Eq.~\eqref{appeq:leading_contribution_permutations}.
We aim for showing that $\mu \in G_m$.
To this end write $\mu(0) = \mu(0 + 0t) = r_0 + k_0t$ as introduced above.
Equation~\eqref{appeq:leading_contribution_permutations} then implies 
$\mu(s) = \left[(r_0 + s)\!\!\mod t\right] + nt=\eta_{r_0}(s) + k_0t$ for all $s \in \{0,1, \ldots, t-1\}$.
That is the value of $\mu(s + 0t)$ is fixed by $\mu(0 + 0t)$.
Similarly the values of $\mu(s + nt)$ are fixed by $\mu(nt) = r_n + k_nt= \eta_{r_n}(0) + k_nt$.
As $\mu$ is a bijection all $k_n$ are pairwise distinct and hence define
$\rho \in S_m$ via $\rho(n) = k_n$.
Combining the above arguments yields $\mu = \left(\rho, \eta_{r_0},\ldots,\eta_{r_{m-1}}\right) \in G_m$ and consequently $G_m$ exhausts all solutions to Eq.~\eqref{appeq:leading_contribution_permutations}.

\subsection{Counting Fixed Points}
\label{App:FixedPoints}

As argued in Sec.~\ref{sec:sff_semiclassics} and Sec.~\ref{sec:sff_moments} the average over the phases, Eq.~\eqref{eq:phase_average}, is determined by the number of fixed points of $\mu= \left(\rho, \eta_{r_0}, \ldots, \eta_{r_{m-1}}\right) \in G_m$. 
It is easy to see, that $x=s + nt \in \{0,1, \ldots, mt-1\}$ is a fixed point of $\mu$ if and only if $\rho(n)=n$ and $r_{n} = 0$.
The latter implies that all $x^\prime = s^\prime + nt$ are fixed points as well and hence fixed points of $\mu$ come in multiples of $t$. 
We aim for counting the number of permutations in $G_m$ with exactly $kt$ fixed points.
To this end denote by $B_l$ the number of permutations in $S_m$ with exactly $l$ 
fixed points and by $C_k^l(t)$ the number of permutations $\mu = \left(\rho, 
\eta_{r_0}, \ldots, \eta_{r_{m-1}}\right)$ with $kt$ fixed points under the 
constraint that $\rho \in S_m$ has exactly $l$ fixed points.
Using this definitions allows for writing
\begin{align}
    A_k(t) = \sum_{l=k}^{m}B_lC^l_k(t)
    \label{appeq:A_kt}
\end{align}
and we are left with computing $C_k^l(t)$ and $B_l$.
The latter is the well known combinatorical problem of counting so called 
partial derangements. 
That is, $B_l$ is the number of possibilities to rearrange $m$ objects while keeping $l$ of them in their original place.
It is given by
\begin{align}
B_l = \binom{m}{l}!\left(m-l\right),
\label{appeg:B_l}
\end{align}
where $!x$ denotes the subfactorial.
To compute $C^l_k(t)$ assume that for $\mu =  \left(\rho, \eta_{r_0}, \ldots, \eta_{r_{m-1}}\right)$ the permutation $\rho \in S_m$ has exactly $l$ fixed points. 
For each $n \in \{0,1,\ldots,m-1\}$ which is not a fixed point of $\rho$ there are $t^{m-l}$ choices for the $r_n$.
Among the $l$ fixed points $n$ one needs to choose $k$ with $r_n=0$ for which there are $\binom{l}{k}$ choices.
This guarantees, that $\mu$ has at least $kt$ fixed points.
In order for $\mu$ to have no additional fixed points $r_n$ must be different from zero for the remaining $l-k$ fixed points of $\rho$.
There are $(t-1)^{l-k}$ possible choices.
Combining the above counting arguments yields 
\begin{align}
    C^l_k(t) = \binom{l}{k} t^{m-l}(t-1)^{l-k}.
    \label{appeq:C_kl}
\end{align}
Inserting  $B_l$, Eq.~\eqref{appeg:B_l}, and $C^l_k(t)$, Eq.~\eqref{appeq:C_kl} into Eq.~\eqref{appeq:A_kt} proves the validity of Eq.~\eqref{eq:A_kt}.

\section{Derivation of the Transition Parameter}
\label{App:PerturbationTheoryTransitionParameter}

In Sec.~\ref{sec:perturbation_theory} we introduced the transition parameter $\Lambda$.
In the following we provide a detailed computation leading to Eq.~\eqref{eq:transition_parameter}.
For convenience we repeat the definition \cite{SriTomLakKetBae2016}
\begin{equation}
 \Lambda = \frac{\epsilon^2 \nu^2}{D^2},
\end{equation}
where $D = \frac{2 \pi}{N^L}$ is the mean level spacing of $\U$.
By $\nu^2$ we denote the 
mean value of the modulus squared of the off diagonal entries of the perturbation operator $V$ in the product
eigenbasis of the non-interacting system. 
In leading order in $\epsilon$, computing the product $\tilde{\nu}^2 = \epsilon^2\nu^2$ from this definition is equivalent to computing the mean value of the modulus squared of the off diagonal entries of the interaction $\U_{\uc}(\epsilon)$.
The difference in the computation is essentially the order in which one does the RMTE average and the expansion of the exponential $\U_{\uc}(\epsilon) = \exp(\ui \epsilon V)$.
The resulting transition parameters are related by Eqs.~\eqref{eq:transition_parameter}~and~\eqref{eq:lambda_vs_gamma}.

To compute $\tilde{\nu}^2$ let $W_{IJ}$ denote the unitary matrix which diagonalizes the uncoupled system.
In this basis the matrix elements of the interaction $\U_{\uc}$ read
\begin{align}
 z_{I M} &= \sum_{J,K}W_{KI}^* \left( \U_\uc \right)_{KJ} W_{JM} 
 \label{app_eq:matrix_elements}
\end{align}
where the sums run over all $N^L$ product states. 
This yields
\begin{align}
 \tilde{\nu}^2 &= \Big\langle\frac{\sum_{I,M} |z_{IM}|^2 - \sum_{I} |z_{II}|^2}
            {N^{2 L} - N^L}\Big\rangle \\
       &= \frac{N^L - \langle d \rangle }{N^{2L} - N^L},
\end{align}
where the brackets denote the average over the extended RMTE.
In the first line we wrote the mean over off-diagonal elements as the 
sum over all matrix elements and subtracted the diagonal part. The 
second line follows from the normalization of the rows (or columns) of unitary matrices.
The $N^L$ normalized rows yield the first 
term and we are left with computing the average of the diagonal term $d$ only. 
Using Eq.~\eqref{app_eq:matrix_elements} the latter is written 
explicitly as
\begin{align}
 d &= \sum_{K,M} \left(\U_\uc \right)_{KK} \left( \U_\uc^* \right)_{MM}
        \sum_{I} \left| W_{IK} \right|^2 \left| W_{IM} \right|^2.
\end{align}
We now perform the average over the $L$ independent subsytems.
The eigenbasis $W$ of the unperturbed system $\U_0 = \U_1 \otimes \ldots \otimes 
\U_L$, is itself a CUE matrix of the same structure $W = w_1 \otimes \ldots 
\otimes w_L$, and hence its matrix elements can be written as products
\begin{align}
 \left| W_{IK} \right|^2 &= |w_{i_1 k_1}|^2 \cdots |w_{i_L k_L}|^2.
\end{align}
Due to this factorization, we can restrict to expressions of the form 
\begin{align}
 t_j &= |w_{i_jk_j}|^2 |w_{i_jm_j}|^2
\end{align}
when performing the average over the $j$-th subsystem.
The latter corresponds to the Haar average over the unitaries $w_j$ and yields 
 the second moment of the CUE(N) which reads \cite{PucMis2017}
\begin{align}
 \langle t_j \rangle &= \frac{\delta_{k_j m_j} + 1}{N(N+1)}.
\end{align}
Hence we obtain for the averaged diagonal contribution
\begin{align} 
 \langle d \rangle  & = 
         \sum_{K,M} \Big\langle\left(U_\uc \right)_{KK} \left( U_\uc^*\right)_{MM}\Big\rangle_{\xi} \nonumber 
        \\
        &\qquad \times N^L\left[\frac{\delta_{k_1 m_1} + 1}{N(N+1)} \cdots
                       \frac{\delta_{k_L m_L} + 1}{N(N+1)}\right] 
        \label{app_eq:average_diagonal2}
\end{align}
with only the average over the phases $\xi_I$ left.

To perform this remaining average, we note, that each pair of states $K,M$ contributes to 
the above sum as 
\begin{align}
\frac{1}{(N+1)^L}\Big\langle \exp\left(\ui  \epsilon \left[\xi_K - \xi_M\right] \right) \Big\rangle_{\!\xi} 2^{\Delta_{K,M}}, 
\end{align}
where $\Delta_{K,M} = \sum_{i=1}^L \delta_{k_im_i}$ is the number of subsystems for which the factors of the product states labeled by $K$ and $M$ agree.
Similar as described in the main text in Sec.~\ref{sec:sff_semiclassics} the average over the phases
reads
\begin{align}
    \Big\langle \exp\left(\ui  \epsilon \left[\xi_K - \xi_M\right] \right) \Big\rangle_{\!\xi} = \left(1 - \delta_{KM}\right)|\chi(\epsilon)|^2 + \delta_{KM} 
\end{align}
due to the $\xi_I$ being i.i.d. random variables and where again $\delta_{KM}$ is understood element wise.
As the average over the phases depends on $\Delta_{K,M}$ only, evaluating the sum over $K,M$ in Eq.~\eqref{app_eq:average_diagonal2} can be achieved by counting the number of pairs $K,M$ which yield the same value of $\Delta_{K,M}$.
To this end, let $D_j$ denote the number of pairs for which $\Delta_{K,M}=j$.
This allows for writing 
\begin{align}
\langle d \rangle & = \frac{D_L2^L}{(N+1)^L} + \frac{|\chi(\epsilon)|^2}{(N+1)^L}\sum_{j=0}^{L-1}D_j 2^j .
\label{app_eq:average_diagonal}
\end{align}
A standard combinatorical argument gives 
\begin{align}
D_j  = N^L\binom{L}{j}(N-1)^{L-j}
\end{align}
and the sum in Eq.~\eqref{app_eq:average_diagonal} can be written as 
\begin{align}
\sum_{j=0}^{L-1}D_j 2^j = N^L(N+1)^L - (2N)^L
\end{align}
by means of the binomial theorem.
Inserting this into Eq.~\eqref{app_eq:average_diagonal} and the resulting expression for $\langle d \rangle$ into Eq.~\eqref{app_eq:average_diagonal2} we obtain
\begin{align}
\tilde{\nu}^2 = \frac{1}{\left(N^L - 1\right)}\left(1 - \left(\frac{2}{N+1}\right)^L\right)\left(1 - |\chi(\epsilon)|^2\right).
\end{align}
Consequently, this gives the transition parameter as 
\begin{align}
\Lambda = \frac{N^L}{4\pi^2}\left(1 - |\chi(\epsilon)|^2\right),
\end{align}
where we neglected subleading terms in $1/N$ and which corresponds to Eq.~\eqref{eq:transition_parameter} in the main text.

\section{Calculation of the Perturbation Integral}
\label{App:PerturbationTheoryIntegral}

The integral~\eqref{eq:perturb_int} can be solved 
using multiple substitutions. First, we 
replace $z = s / \sqrt{\Lambda}$ and thus unravel the integration variables 
and system parameters. This motivates the introduction of the rescaled time $\tau_\text{pert} 
= D \sqrt{\Lambda} t$, Eq.~\eqref{eq:tau_pert}.
The integral then reads 
\begin{align}
  \mathcal{I} &=\sqrt{\Lambda} \int_{-\infty}^\infty\!\! \ud z\int_0^\infty \!\!\ud w \, \cos(\tau_\text{pert} 
\sqrt{z^2 + 
4 w}) \ue^{-w}.
\end{align}
By further substituting $w = y^2$ and introducing polar coordinates $r = z^2 + y^2$ and $y = r \sin(\theta)$ we 
obtain
\begin{align}
     \mathcal{I} &= \frac{\sqrt{\Lambda}}{2} \int_{-\infty}^{\infty} \ud z 
\int_{0}^{\infty} 
 \ud y \, y \cos(\tau_\text{pert} \sqrt{z^2 + y^2}) \ue^{-y^2/4}, \\
 \begin{split}
 &= \frac{\sqrt{\Lambda}}{2} \int_0^\infty \ud r \int_{0}^\pi \ud \theta \,
 \left( r^2 \sin(\theta) \ue^{-r^2 \sin(\theta)^2/4} \right. \\
  &\qquad\qquad\qquad\qquad  \left. \times \frac{\ue^{-
\ui\tau_\text{pert} r} + \ue^{\ui \tau_\text{pert} r}}{2} \right)
\end{split}\\
\begin{split}
 &= \frac{\sqrt{\Lambda}}{4} \int_{-\infty}^\infty \ud r \int_{0}^\pi 
\ud \theta \, \left( r^2 \sin(\theta) \right. \\
        &\qquad\qquad \left. \times \ue^{-r^2 \sin(\theta)^2/4}
\ue^{\ui\tau_\text{pert} r}  \right).
\end{split}
\end{align}
The integral over $r$ takes the form of a Fourier transform and yields
\begin{align}
\begin{split}
   \mathcal{I} &= - \sqrt{\pi \Lambda} \int_0^\pi \ud  \theta \,  
    \ue^{-\tau_\text{pert}/\sin^2(\theta)} \times \\
    &\qquad\qquad \times \left(\frac{2 
\tau_\text{pert}^2}{\sin^4(\theta)} - \frac{1}{\sin^2(\theta)}\right)
\end{split}\\
 &= -2 \pi \sqrt{\Lambda} \tau_\text{pert} \ue^{-\tau_\text{pert}^2}.
\end{align}
This is the expression~\eqref{eq:sff_perturbativ_integral} as stated in the main text.

\bibliography{paper_sff_rmte}

\end{document}